\documentclass[]{agujournal2}
\draftfalse
\usepackage{graphicx}
\usepackage{enumitem}
\usepackage{color}
\usepackage{bm}
\usepackage[nolist]{acronym}
\usepackage{gensymb}
\begin{document}
\title{Recent progress and review of issues related to Physics Dynamics Coupling in geophysical models}
\authors{Markus Gross\affil{1},
Hui Wan\affil{2},
Philip J. Rasch\affil{2},
Peter M. Caldwell \affil{3},
David L. Williamson\affil{4},
Daniel Klocke\affil{5},
Christiane Jablonowski\affil{6},
Diana R. Thatcher\affil{6},
Nigel Wood\affil{7},
Mike Cullen\affil{7},
Bob Beare\affil{8},
Martin Willett\affil{7},
Florian Lemari\'{e}\affil{9},
Eric Blayo\affil{9},
Sylvie Malardel\affil{10},
Piet Termonia\affil{11},
Almut Gassmann\affil{12},
Peter H. Lauritzen\affil{4},
Hans Johansen\affil{13},
Colin M. Zarzycki\affil{4},
Koichi Sakaguchi\affil{2},
and Ruby Leung\affil{2}
}
\begin{acronym}
 \acro{ENDGame}{\emph{E}ven \emph{N}ever \emph{D}ynamics for \emph{G}eneral \emph{a}tmospheric \emph{m}odelling of the \emph{e}nvironment}
 \acro{OA}{Ocean-Atmosphere}
 \acro{ALADIN}{\emph{A}\/ire \emph{L}\/imit\'ee \emph{A}\/daptation dynamique \emph{D}\/\'eveloppement \emph{I}\/nter\emph{N}\/ational}
 \acro{ALARO}{\emph{Ala}\/din A\emph{ro}\/me}
 \acro{CAM}{Community Atmosphere Model}
 \acro{GCM}{General Circulation Model}
 \acro{SL}{Semi Lagrangian}
 \acro{IFS}{Integrated Forecasting System}
 \acro{ECMWF}{European Centre for Medium-Range Weather Forecasts}
 \acro{ECHAM}{European Centre - Hamburg model}
 \acro{ECHAM5}{\ac{ECHAM} Version 5}
 \acro{CO2}[CO$_{2}$]{carbon di-oxide}
 \acro{SLAVEPP}{Semi-Lagrangian Averaging of Physical Parameterizations}
 \acro{CAPE}{Convective Available Potential Energy}
 \acro{ECHAM-HAM}{}
 \acro{NCAR}{National Center for Atmospheric Research}
 \acro{ITCZ}{Inter-Tropical Convergence Zone}
 \acro{EDMF}{Eddy Diffusivity Mass Flux}
 \acro{CLUBB}{Cloud Layers Unified by Binormals}
 \acro{NOGAPS}{Navy Operational Global Atmospheric Prediction System}
 \acro{QG}{quasigeostrophic}
 \acro{EMBRACE}{Earth system model bias reduction and assessing abrupt climate change}
 \acro{UK}{United Kingdom}
 \acro{UTC}{}
 \acro{Ro}{Rossby number}
 \acro{HPE}{hydrostatic primitive equation}
 \acro{SST}{sea surface temperature}
 \acro{MITC}{ Moist Idealized Test Case }
 \acro{NWP}{numerical weather prediction}
 \acro{RMS}{}
 \acro{FV}{Finite-Volume}
 \acro{SE}{Spectral Element}
 \acro{UW}{University of Washington}
 \acro{PBL}{ planetary boundary layer }
 \acro{APE}{Aqua-Planet Experiment}
 \acro{WRF}{Weather Research and Forecasting}
 \acro{ROMS}{Regional Ocean Modeling System}
 \acro{SCM}{Single Column Model}
 \acro{LES}{Large Eddy Simulation}
 \acro{PDC}{Physics Dynamics Coupling}
 \acro{TKE}{subgrid-scale kinetic energy}
 \acro{LAM}{Limited Area Model}
 \acro{CSRM}{Cloud-System-Resolving Model}
 \acro{3MT}{Modular Multiscale Microphysics and Transport scheme}
 \acro{KE}{kinetic energy}
 \acro{GLL}{Gauss-Lobatto-Legendre}
 \acro{MPAS-A}{Model for Prediction Across Scales - Atmosphere}
 \acro{AMIP}{Atmospheric Model Intercomparison Project}
\acro{CCM}{\ac{NCAR} Community Climate Model}
\acro{CCM3}{\ac{CCM} Version 3}
\acro{VR}{variable-resolution}
\acro{QU}{quasi-uniform}
\acro{CFL}{Courant--Friedrichs--Lewy}
\acro{SCVT}{spherical controidal Voronoi tessellations}
 \end{acronym}
\affiliation{1}{Departamento de Oceanograf\'{i}a F\'{i}sica, Carretera Ensenada-Tijuana 3918, Ensenada BC 22860, M\'{e}xico.}
\affiliation{2}{Pacific Northwest National Laboratory, 902 Battelle Boulevard, Richland, WA 99354, USA.}
\affiliation{3}{Physical and Life Sciences Directorate, Lawrence Livermore National Laboratory, 7000 East Avenue, Livermore, CA 94550, USA}
\affiliation{4}{National Center for Atmospheric Research, P.O. Box 3000, Boulder, CO 80307-3000, USA.}
\affiliation{5}{Hans Ertel Center for Weather Research, Deutscher Wetterdienst (DWD), Germany}
\affiliation{6}{University of Michigan, Department of Climate and Space Sciences and Engineering, 2455 Hayward St., Ann Arbor, MI 48109, USA.}
\affiliation{7}{Met Office, FitzRoy Road, Exeter, EX1 3PB, United Kingdom.}
\affiliation{8}{CEMPS, Exeter University, Prince of Wales Road, Exeter, EX4 4SB, UK.}
\affiliation{9}{INRIA, Univ. Grenoble-Alpes, LJK, CNRS, Grenoble F-38000, France.}
\affiliation{10}{ECMWF, Shinfield Park, Reading, RG2 9AX, UK.}
\affiliation{11}{Royal Meteorological Institute of Belgium, Ringlaan 3, Avenue Circulaire, B-1180 Brussels, Belgium.}
\affiliation{12}{IAP K\"{u}hlungsborn, Leibniz-Institut f\"{u}r Atmosph\"{a}renphysik e.V. an der Universit\"{a}t Rostock, Schlossstra{\ss}e 6, 18225 K\"{u}hlungsborn, Germany}
\affiliation{13}{Applied Numerical Algorithms Group, Lawrence Berkeley National Lab, 1 Cyclotron Road, Berkeley, CA 94720, USA}
\correspondingauthor{M Gross}{mgross@cicese.mx}
\begin{abstract}

Geophysical models of the atmosphere and ocean invariably involve parameterizations. These represent two distinct areas: Subgrid processes that the model cannot resolve, and diabatic sources in the equations, due to radiation for example. Hence, coupling between these physics parameterizations and the resolved fluid dynamics and also between the dynamics of the air and water, is necessary. In this paper weather and climate models are used to illustrate the problems. Nevertheless the same applies to other geophysical models. This coupling is an important aspect of geophysical models. However, often model development is strictly segregated into either physics or dynamics.
As a consequence, this area has many unanswered questions. Recent developments in the design of dynamical cores, extended process physics and predicted future changes of the computational infrastructure are increasing complexity.
This paper reviews the state-of-the-art of the physics-dynamics coupling in geophysical models, surveys the analysis techniques, and illustrates open questions in this field.
This paper focuses on two objectives: To illustrate the phenomenology of the coupling problem with references to examples in the literature and to show how the problem can be analysed. Proposals are made on how to advance the understanding and upcoming challenges with emerging modeling strategies. This paper is of interest to model developers who aim to improve the models and have to make choices on and test new implementations, to users who have to understand choices presented to them and finally users of outputs, who have to distinguish physical features from numerical problems in the model data.
\end{abstract}

%
%

\section{ Introduction }
\begin{figure*}
\centering
\includegraphics{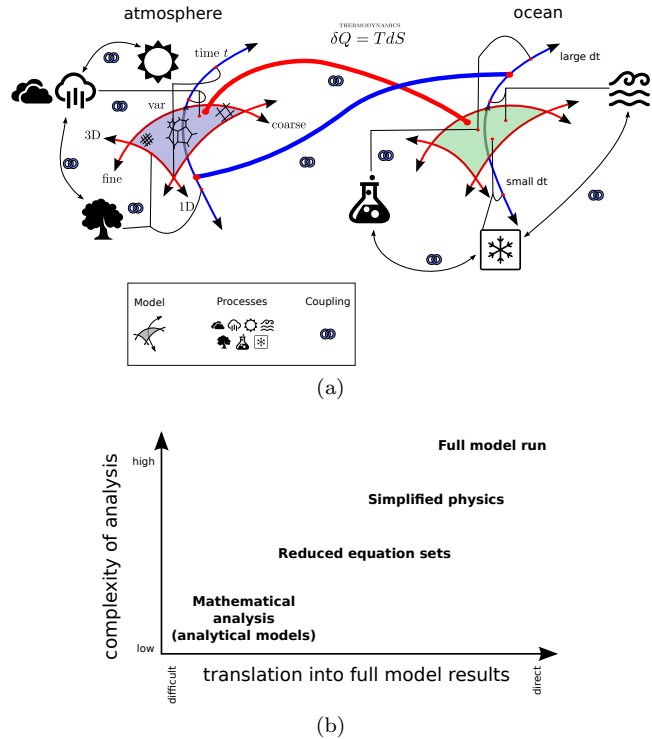}
 \caption{Schematic representation of \ac{PDC}. For simplicity panel a) shows only two models, an ocean model and an atmosphere model. Both of these have spatial scales (here indicated by the plane with red lines) and temporal scales (indicated by the blue axis). These are coupled (thick lines), i.e. one domain in the spatial plane maps into the spatial plane of the other model (thick red line) and similarly in the temporal axis (thick blue line). In the spatial plane aspects such as grid type, fixed versus variable resolution, one dimensional vs three dimensional and fine versus coarse are shown as some of the aspects of the spatial resolution that can vary in between models and does not necessarily have a straight forward mapping. Then each of these models has its own ecosystem of parameterizations (an arbitrary set of processes was chosen here for illustration only), which interact with the model and themselves - via coupling. These parameterizations as well occupy potentially - or almost certainly - different areas on the spatial plane and temporal axis.  All of this exists in front of a background problem of thermodynamics, which ultimately governs them all (or is ought to, anyhow).
Panel b) shows the four tier scheme of investigation, ranging from (by necessity) abstract analysis, via reduced equation sets (with less necessity for abstraction) to simplified physics tests and finally full model runs. The complexity of the analysis increases from one to the other. The manner in which the results and conclusions from the experimentation can inform the production runs ranges from ``difficult'', i.e. results are expected in the form of guidance or informing a choice that needs to be made in the design phase, to ``direct'', meaning that a benefit can be demonstrated straight away by producing an improved forecast.}
  \label{fig:PDC}
 \end{figure*}

In the context of this publication geophysical models are weather, climate, and Earth system models that describe fluid dynamics of the oceans and the atmosphere as well as their interactions with various physical, chemical, and biogeochemical processes that occur within those fluids or at the Earth's surface. The aim of such a geophysical model is to predict a spatially and temporally discrete representation of the true solution. This true solution is defined by a set of equations describing the physics (e.g., balances of momentum, energy and mass) and chemistry (and possibly even the biogeochemistry) of the  geophysical system. Discrete approximations, in space and time, to these equations are necessary in order to numerically solve these equations using a computer,  to produce simulations approximating the original physical system in the form of a space-time average of the governing equations.
 Spatial and temporal discretizations are two distinct yet related aspects, and are symbolically shown in Figure~\ref{fig:PDC}a by the curved surfaces and the arrows perpendicular to the surfaces, respectively. Considering the finite resolutions that are practically affordable in terms of computational cost, some component models are further divided into sub components representing processes (phenomena) that are resolved or unresolved (under-resolved). In the atmosphere and ocean models, the sub components that describe the resolved fluid dynamics are commonly known as the dynamical cores or simply ``dynamics", while the representation of unresolved or under-resolved processes is referred to as the subgrid-scale parameterization or simply ``physics", which operate on time and space scales much below the model resolution.

In this context \ac{PDC} is defined as the formulation and implementation of the coupling between any two (or more) physical components of the modelling system under consideration. The term physical component is used to represent any of: an individual physical parameterization; a collection of such parametrizations; the dynamical core (for example, of the atmosphere or the ocean); or a modelling subsystem (such as the atmosphere and ocean models in an Earth System Model). The formulation and implementation of the coupling should address the following issues: 
\begin{itemize}
\item  the compatibility of the thermodynamic formulation between components;
\item  the discrete representation of the interaction between components that represent a possibly vast (and vastly different) range of time and space scales;
\item  the possible use of different resolutions between components (including variable versus fixed resolutions);
\item  and different spatial and temporal discretizations of the governing equations (for example spectral versus grid point versus finite element).
\end{itemize}
Thererfore, as Figure~\ref{fig:PDC}a aims to illustrate, \ac{PDC} is not limited to only the one dimensional interaction between physics and dynamics.
A key challange in the above is the design of space-time integration schemes for the different components that, when combined, reproduce the space-time averaged behaviour of \emph{the whole system} being modelled.


In part due to the very high level of complexity of the real-world system, those models are typically developed, evaluated, and applied in units called component models that correspond to the different target systems, for example the atmosphere, ocean, land, glacier, and sea ice. The schematic shown as Figure~\ref{fig:PDC}a includes only two component models for simplicity: the atmosphere and the ocean. Those components are inherently coupled to each other through the momentum, mass and energy exchanges at their interfaces.

The parameterizations are typically organized by processes, for example cumulus convection and cloud microphysics in the atmosphere, and lateral and vertical mixing in the ocean. Some of these processes are symbolized by clip art icons in Figure~\ref{fig:PDC}a. Different processes can, and do in real models, reside at different locations in the space-time domain. For example the characteristic time scales associated with cloud microphysics and planetary-scale advection are vastly different. 

The wide ranges of spatial and temporal scales that are associated with the different elements in Figure~\ref{fig:PDC}a have naturally resulted in different foci in research and the compartmentalization of the model codes. This compartmentalization and separation is necessary in order to  understand and gain insights into the complex system and to render the model development manageable and traceable.

This, however, leads to what is known as splitting, i.e., evaluating in isolation, in time and or space, the response and feedback of a process to the evolution of model state, assuming the other processes stay unchanged during a certain time interval or that processes are evaluated in a pre-determined hierarchy, sequentially progressing the model state from one process (or family of processes) to the next. While splitting is useful and unavoidable, it can also lead to undesirable features in the numerical solutions since many processes are linked (coupled) to each other. In fact, many interactions are known to exist between the dynamical cores and the parameterizations, between different parameterized processes, and between component models such as the atmosphere and ocean. The multiplicity of timescales and their broad span from microseconds to months means that the time averaging required by the numerical solution is highly non-trivial.

The modeling errors inevitably introduced by the splitting procedure as outlined above  are the core theme of the present paper. This is arguably a key, yet under investigated topic (however the community is starting to embrace the topic \citep{Gross2015}). In the past, the much lower spatial resolution and much simpler model formulation have been the dominant sources of model error. In recent years, however, rapid enhancement of computing capabilities has allowed for substantial increase in model resolution as well as the incorporation of much more comprehensive description of subgrid-scale phenomena. Examples of the latter include the life cycles of atmospheric aerosols and cloud droplets, which involve processes at spatial scales of nanometers to microns. If the coupling between the components does not "transport" sufficient information back and forth from the dynamical core to the physics, then the most accurate scheme in the dynamics alone will not improve the quality of the model simulations as may be otherwise expected. Sufficient information is to be understood as each component being able to utilise the input from other components in a physical meaningful and compatible way. For example, high order dynamics will not be able to reveal its full potential if coupled with low order physics. The low order physics cannot react physically to the high order information from the dynamics. Nevertheless the physics will create forcings which in turn will influence the dynamics significantly.  Thus numerical issues in the parameterized physics and in the process coupling can be bottlenecks in the reduction of overall model error.  Also, in the prediction of the discrete representation (which may be point wise or space-time average over the grid box) of the true solution a subgrid model is required, but the subgrid model can only be formulated by assuming a scale separation, between the resolved and unresolved scale. This scale seperation depends on reolution and becomes more difficult and or questionable as resolution increases (c.f.,  sections~\ref{sec:simplified_eq} and~\ref{sec:grey}). 

  The present paper presents examples and reviews the ongoing efforts in addressing those issues, and discuss an overarching topic of thermodynamic consistency across model components. Also discussed are a hierarchy of analysis methods aiming at furthering the understanding and provide a means of analysis of the intricate interactions, using mathematical analysis, reduced equations, model with simplified physics, and full model analysis.

The remainder of the paper is organized as follows:

\emph{Section~\ref{sec:hist}} reviews the historical development of the field, indicating key publications and advances, analysis and tests performed so far and the evidence available. 
\emph{Sections~\ref{sec:arti}} focuses on issues related to process splitting in the time stepping algorithm, which can strongly affect model behavior in various ways.  Process splitting is commonly used to make the approximation of each process more practical and to obtain an easy-to-maintain modular structure of the model source code. It may be argued that this is a pragmatic and/or convenient choice. Since different processes can produce compensating effects or can compete for resources, large numerical errors can result if processes are allowed to operate in isolation for periods of time longer than the typical time scales of their interactions. Solution convergence with respect to time step size is discussed in section~\ref{sec:time} in the context of the spatial resolution being kept constant. While increasing only the temporal resolution does not guarantee that the solution will converge to the observed atmospheric motions in the real world, it is argued that the convergence analysis can be informative and help to improve the understanding of the process interactions in a model.

The following section (\emph{section~\ref{sec:simplified_eq}}) then proceeds to discuss that while analytically a convergence study (in time and space) will eventually converge towards the solution of the Navier Stokes equations, this would require resolution of the Kolmogorov scale, which is far from current resolutions and computational capability. Therefore, alternatively, an artificial scale separation has to be imposed if convergence is required or desired.

\emph{Section~\ref{sec:simplified_eq}} illustrates methods of evaluating the coupling strategies. It is first necessary to distinguish the parts of the physics that force the system (e.g. radiation) from those (the subgrid model) that
represent subgrid dynamics and thermodynamics and vanish at infinite resolution. Coupling of the subgrid model can be evaluated using a (reduced) continuous equation set. This defines a scale separation and allows a precise definition of the required solution in which the subgrid model plays an important part. In particular, the accurate reproduction of asymptotic limits where subgrid transports play a key role can be checked. Different coupling strategies can thus be analyzed in a less abstract way than the analysis by  \cite{QJ:QJ200613261402}.
The importance of validating subgrid models and coupling strategies against the averaged data from much higher resolution models is illustrated using a convection example.

\emph{Section~\ref{sec:simple}} emphasizes that ideally there would be a standard test procedure and established benchmark results across a whole range of models, with tests that isolate the components while still reflecting the model complexity and hence maintaining relevance. However, so far the design and testing strategy for simplified tests which stress the coupling and provide useful results has proven very difficult indeed. A proposal for idealized testing of the coupling is made and some results from already implemented tests are provided.

\emph{Section~\ref{sec:intra}} focuses on the coupling in between different models, as depicted in Figure~\ref{fig:PDC}. An exchange of information has to occur on both the space plane and the time axis.
Numerical results suggest a correlation between coupling errors and model uncertainties. \ac{OA} coupling requires the careful study of the mathematical and numerical formulation of physical parameterizations. In analogy to section~\ref{sec:simple},  a lack of simplified models and reference test cases  is observed.

\emph{Section~\ref{sec:thermod}} summarizes the state of the current scientific discussion around the thermodynamic compatibility. Even with ideal coupling in the time and space domains, different models are coupled. This has to be done in accordance with the laws of thermodynamics. The dynamics of a model describes exclusively reversible phenomena, whereas the physics of a model describes irreversible phenomena.

\emph{Sections~\ref{sec:grey} and \ref{sec:emerging}} discuss the complexity of the interaction of parameterizations and resolution change. As resolution increases more detail is included in the solution to the governing equations, leaving less for the parameterizations. An example, amongst others, are mesoscale eddies in the oceans. Also some assumptions cease to be valid, such as a single continuity equation in the presence of sea ice. In this paper, first issues relating to uncertainties in the parameterizations themselves as horizontal resolutions approach and pass convection permitting regimes, the gray zone, are defined and illustrated with examples. Limitations of the current models in the gray zones alongside a short review of the progress to date are presented and illustrated using the example of the \ac{ALARO} configuration of the \ac{ALADIN} model \citep{DeTro13}.

Section~\ref{sec:emerging} discuss new and emerging modeling strategies of separating physics and dynamics grids (\ref{sec:emerging_fe}), and how time-stepping/process-splitting (sections~\ref{sec:hist} and~\ref{sec:arti}) and scale-awareness of deep convection (section~\ref{sec:grey}) can interact and pose a challenge to models using spatialy varying horizontal resolution (\ref{sec:varres})

\emph{Section~\ref{sec:conc}} summarizes the contributions from the different areas in the context of the whole picture elaborated on in the individual sections, aims to strengthen the awareness of this aspect in the community and invites the community to resolve and understand the issues jointly.

\section{ Historical review \label{sec:hist}}
The history of \ac{PDC} probably starts with the first \ac{GCM} simulations. In the late 1960s, \cite{Syukuro69}, at the Geophysical Fluid Dynamics Laboratory in Princeton, New Jersey, presented one of the first simulations of the global atmospheric circulation coupled to ocean processes.

There were two models: one for the atmosphere, developed by Manabe, a meteorologist, who developed the atmosphere model; and one for the Ocean, developed by Bryan, an oceanographer with meteorological training. This separation of model domains remains today, nearly 50 years later. Manabe and Bryan joined forces to create a computational system that coupled their models. The winds and rain would contribute to the ocean currents, and the sea-surface temperatures and evaporation contribute to the circulation of the atmosphere. They soon realized that the coupling required much more than just passing on the forcing from one model to the other. Simply computing the fluxes in one model and applying them to the other did not yield a satisfactory outcome.

In the early days the coupling problem was a practical one: how to combine the different time scales and fit the model to the available computing capacity. The challenge was to get a working model.

Over the next twenty years computational resources were increasing at an accelerating rate. This had two implications. First, there was the capacity to run more and more coupled models and to increase the complexity. And second, with more data available it became possible to study the interaction, the coupling in more detail. The emphasis shifted from getting it running in the first place to investigating and contrasting different strategies and formulations. The literature details a long series of investigations and reports of problems faced when coupling the dynamical systems to their forcings.

\cite{Lander1997} investigated and discussed the scales generated by an atmospheric model and argued that not all of them should be utilized in the physical parameterizations, because they are artifacts of the solution procedure, and not part of the solution itself. If the scales are close to the truncation limit and the model is not strongly damped there is significant noise present in the solution. If, partly in response to the noise,  the model is damped then it is most strongly so near the truncation limit. Also the discretization error, which is always present, is most significant here. They therefore suggest that these ``unbelievable'' scales should not be used in the parameterizations and recommend the use of a coarser grid to evaluate the forcing from the parameterizations. The non-linear character of some physical parameterizations and their sensitivity to small perturbations can otherwise quickly lead to the growth of noise, rather than a correct approximation of the physics of the system (cf. sections~\ref{sec:grey} and \ref{sec:emerging_fe}). Interestingly recent work carried out at the \ac{ECMWF} goes in the opposite direction, i.e. using higher spatial resolution for the parameterizations. Their results clearly show benefit from doing this, as further elaborated in section~\ref{sec:grey}.

\cite{Caya1998} investigated temporal aspects of the coupling. They discuss the effect of ``splitting'', i.e. applying the parameterizations subsequently to time stepping the dynamics, in combination with long time steps (15 min), such as admissible by \ac{SL} models. They argue that the splitting error can become unacceptably large.

Probably one of the first studies utilizing the full model and varying coupling parameters systematically is \cite{TELA:TELA0009}. Here the grid of the physical parameterizations and scale of the external surface forcing are held fixed while
the horizontal resolution of the dynamical core is increased. This
is shown to aid the convergence of tropical Hadley circulation, with
increasing dynamical resolution, however it does not converge if the
physics grid is not held constant. This is attributed to the forcing
of smaller scales from the physics, indicating that the parameterizations
at the coarser grid do not include their own forcings from the finer
scale, i.e. missing processes.

Focusing on the vertical and in particular the resolution of the physical parameterizations, \cite{TELA:TELA394} analyzes the full model response to a refinement of the vertical physics grid. It was found that this benefits fields
which are computed directly in the physical parameterizations, and in
the vertical structure of the relative humidity and mass stream function, in line with the \cite{TELA:TELA0009} result, i.e. resolving some of the processes that are not captured by the parameterizations at coarse resolutions.

\cite{tm274} shows that the model performance can be improved by grouping certain parameterizations together and using predictors to improve the input from the dynamics into the parameterizations.

 The suite of parameterizations is split into two groups. One to be evaluated at the arrival point and the other at the departure point. When compared with a simpler fractional stepping (or sequential or time split scheme) the following benefits are observed: second order accuracy, increase in stability, reduction of the time step dependence and numerical noise, improved mass conservation, more accurate forecasts with respect to the root mean square error and anomaly correlations and improved tropical cyclone tracks.

It becomes apparent that there are several options and clearly, some options are better than others. It is not always feasible to construct a new coupling scheme from scratch, implement it and test it in fully operational forecast mode in order to determine if it is better or not. Some form of analysis would be desirable, to test ideas and evaluate potential performance improvements.

Extending the framework presented by \cite{Caya1998},  both in complexity of the sample problems as well as the coupling mechanisms, \cite{QJ:QJ200212858611} and \cite{Staniforth2002} analyze the explicit, implicit, split-implicit and symmetrized split-implicit coupling.  They
highlight that the stability of the explicit coupling is very restrictive
for fast damping processes, such as vertical diffusion in the boundary
layer at high resolution, thus rendering the explicit coupling computationally
inefficient for practical applications. The authors show that this can be addressed by
using implicit coupling, however it leads to ``a highly nonlinear
and computationally difficult and expensive problem to solve''.
The split implicit coupling addresses this but reduces the accuracy.

Back to full model analysis, \cite{Williamson2002} reports statistically
relevant differences when comparing time-split (sequential) and process-split (parallel) couplings to a simulation with the original version of the \ac{CCM3}. However, owing partly to the small time step used, these differences were small, highlighting the difficulty in clearly differentiating better from worse coupling mechanisms using the full model output alone. See also section~\ref{sec:arti}.

\cite{QJ:QJ200312958915} present a predictor corrector scheme that
can give some of the advantages of a fully-implicit scheme and show
that the use of more than one physics evaluation per time step significantly
improves the accuracy in a model problem. 
 An attempt
is made to classify slow and fast processes. Using the predictor scheme
short-time variability is reduced and a transfer from convective
to dynamic precipitation observed in consequence. 
In what is possibly so far the most convincing demonstration on what difference the temporal coupling can make on a forecast,  \cite{Beljaarsetal}
argue that, for the \ac{ECMWF}  \ac{IFS}, sequential splitting (tendencies
of the explicit processes are computed first and are used as input
to the subsequent implicit fast process) is preferable over parallel
splitting (tendencies of all the parameterized processes are computed
independently of each other) for problems with multiple time scales,
because a balance between processes is obtained during the time integration. 
See also section~\ref{sec:arti}.

In an analytically tractable framework - as mentioned above - this practical demonstration of the benefits of the sequential splitting is followed up by \cite{Dubal2004,Dubal2005,QJ:QJ200613261402}, using mathematical analysis. They
conclude that while some advantages exist for parallel splitting
over sequential splitting (e.g., parallel computation and not requiring
an ordering of physical processes), the sequential-split methods are
more flexible when it comes to eliminating splitting errors.

The issue of ongoing non-convergence of model results is highlighted in  \cite{TELA:TELA339}.
Analyzing convergence runs with resolution varying from T42 to T340 truncation and
40 to 5 minutes time step, convergence is observed in larger scales of the zonal
average equatorial precipitation and equatorial wave propagation.
However, a non-convergent mass shift from polar to equatorial regions
and a zonal average cloud fraction decrease was observed. In general,
the simulations show a sensitivity to the parameterizations time step
as well as to the horizontal resolution. Even when the time step is
fixed, global averages do not converge with increasing resolution
for all fields. For example, there is no indication that either precipitable
water or precipitation converges with increasing resolution. This renders the analysis of the coupling more difficult as it is not immediately obvious how to generate a reference solution that can be used to test for coupling errors, using full model runs. The problem of attribution of errors has been more recently investigated in \cite{gmd-6-861-2013}. Only relatively recently has the importance of the coupling in its own right been recognized and efforts to address these issues on a multi disciplinary level are underway \citep{Gross2015}.

This brief - and by no means comprehensive - review is meant to illustrate the vast array of considerations made in the context of coupling models and their parameterizations. Furthermore the difficulties faced when attempting to analyze the impacts and consequences and when designing new coupling algorithms and, indeed, parameterizations, is illustrated.
The remainder of the present publication will illuminate the problem from different directions and highlight current progress in the respective areas, starting with splitting of processes in the discrete model.

\section{ Time stepping errors introduced by splitting }
\label{sec:arti}
Weather and climate models rely on discretizing time and space dimensions in order to make calculations computationally affordable. 
Numerical errors from spatial and temporal discretization can be closely related in some situations 
but more straightforward to separate in other cases.
In this section, the focus is exclusively on time discretization by discussing model behaviors 
with fixed spatial resolution and varied time steps.

\subsection{Impact of time stepping errors}
\label{sec:dt_intro}

\begin{figure} \centering
\includegraphics[width=6cm]{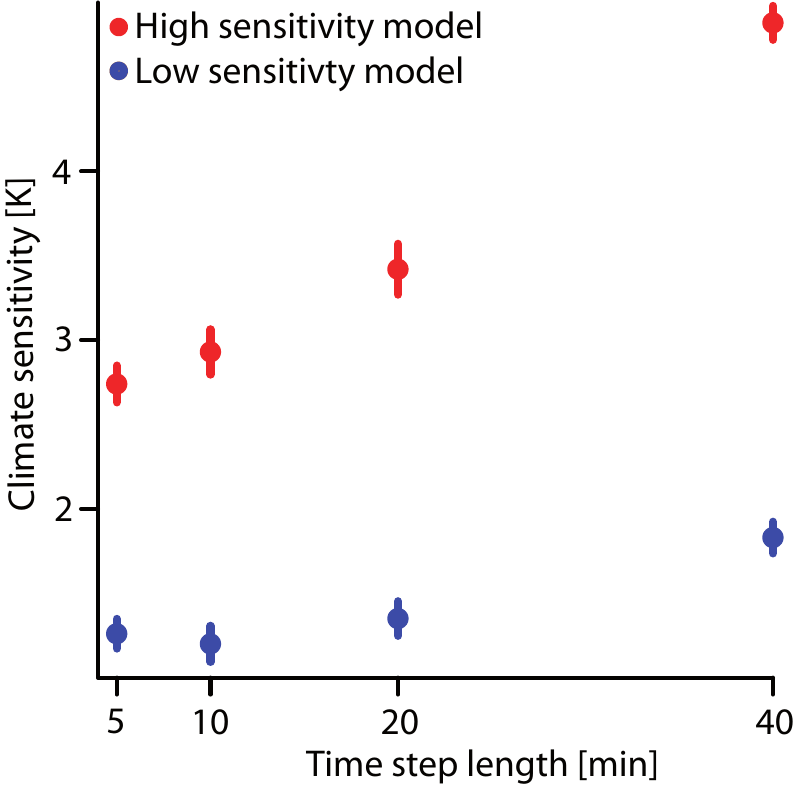}
\caption{
Equilibrium climate sensitivity in \ac{ECHAM5} \citep{roeckner:2003,roeckner:2006} slab-ocean simulations at T31L19 resolution. Red and blue markers indicate high- and low-sensitivity models which differ only in a few uncertain parameters in the physics parameterizations \citep{klocke:2011}. For each time step size listed on the x-axis, the model's climate sensitivity is computed as the difference between $10$~year present-day simulations and the last 10 yrs of a 50 yr doubled \ac{CO2} experiment. Error bars indicate inter-annual variability of global- and annual-mean surface temperature. 
\label{fig:echam_timestep_sensitivity} }
\end{figure}

Time Step size can have a large impact on the behavior of weather and climate models. 
For example, in one version of the \ac{ECHAM5} climate model 
\citep{roeckner:2003,roeckner:2006}, 
the equilibrium climate sensitivity 
(i.e., the global-average equilibrium surface temperature change in response to doubling \ac{CO2})
was found to vary by a factor of two when the model's time step size was varied between 5~min and 40~min
(Figure~\ref{fig:echam_timestep_sensitivity}).
While solution sensitivity to time step size is not at all surprising from a mathematical perspective,
such large discrepancies are undesirable numerical artifacts 
for model users who assume the models reflect
the state-of-the-art understanding of the workings of the real-world system.
In practice, it might be possible to ``tuned away'' the time step sensitivity by using different parameter values 
for different step sizes; however, there exists the danger that such tuning might result in error compensation
that cannot be guaranteed for other applications (e.g., simulations under different forcing scenarios).
To improve the credibility of future climate projections,
it would be useful to revise the model and reduce the sensitivity to time step 
so as to provide the confidence that results from the numerical models are reasonably 
accurate solutions of the underlying continuous physics equations.

Strong sensitivities to model time step have been seen in other models as well. For instance, \citet{wan:2014} showed that clouds and precipitation simulated by the \ac{CAM} version 5 changes substantially when the model time step is reduced from 30~min (the default value) to 4~min. \citet{zhang:2012} found that the impact of swapping aerosol nucleation parameterizations on sulfuric acid gas and aerosol concentrations was overwhelmed by the effect of changing the time stepping scheme used for solving the sulfuric acid gas equation in the aerosol-climate model \ac{ECHAM-HAM}. For ECMWF's weather forecast model IFS, \citet{Beljaarsetal} showed that revising the numerical coupling between the dynamical core and turbulent momentum diffusion can substantially improve the 24 hour forecast of $10$~m wind speed when using a $60$~min time step (which was the operational value at the time). \citet{Williamson2002} mentioned that when the splitting method within the parameterization suite was modified, \ac{CCM3} produced a climate equilibrium that was substantially different from the default model in some small contiguous areas. In other areas, the climates were similar, but the balances producing them were very different. 
Most of the studies cited above and the additional examples mentioned below indicate that \emph{it is often the combination of coupling between processes and long time steps which cause time stepping problems in contemporary models}.
The remainder of this section is focused on coupling issues, though it is acknowledged that long time steps can cause problems within individual processes as well.

\subsection{Splitting in the solution procedure}

In order to facilitate the development, maintenance and practicality of numerical algorithms and model source code, parameterizations in weather and climate models are typically organized as separate modules for different processes. Here processes refer to individual physical phenomena
such as cloud droplet formation or turbulent transport of chemical tracers. These may or may not be implemented in individual parameterizations. The process coupling discussed in this section includes the connection between different parameterizations, the connection between a parameterization and the host \ac{GCM}, or the connection between different physical phenomena within an individual parameterization. Splitting is employed to evaluate the tendency terms for each process and to combine their effects to advance the discrete solution in time. The two most popular methods of splitting in operational models are parallel splitting (computing all process tendencies from the same model state, then using the sum of tendencies to march forward) and sequential splitting (computing a tendency, then either passing it together with the original model state to the next process or updating the model state and passing the new state to the next process). \citet{Beljaarsetal} advocate sequential splitting with processes ordered from slowest to fastest in order to allow processes to feed and balance each other within each model step. It is worth noting that the benefits of sequential splitting depend on what information from already-calculated processes is used in subsequent process calculations. \ac{IFS} uses both state information \emph{and} tendencies from previous processes in some subsequent process calculations (hereafter referred to as sequential-tendency splitting), meaning that the processes see the tendencies of some of the previous process, but the model state is updated at the end of the time step. \ac{CAM} physics simply updates the model state whenever a new tendency is available (hereafter sequential-update splitting). Since sequential-tendency splitting shares more information than sequential-update splitting or parallel splitting, it is unsurprising that it performs better. More sophisticated coupling has also been shown to be beneficial for very specific processes. For example, in the \ac{SLAVEPP} algorithm of \citet{tm274}, the tendencies are evaluated at both the departure and arrival points of the semi Lagrangian trajectory and then averaged. A predictor-corrector scheme is used to connect convective and stratiform clouds with the rest of the model. Below only examples of coupling problems related to long time steps involving sequentially-split models are considered, because this is the prevailing configuration of global models.

\subsection{Issues with splitting}
\label{sec:splittingissues}

One way that splitting causes errors is by skewing the competition for resources (e.g., cloud water, energy, or  \ac{CAPE}) between processes. For example, convective instability can be removed by shallow convection, deep convection, or resolved-scale heating-induced motions. \citet{QJ:QJ1992} provides an example of competition for resources in a sequential-update split model. There it is noted that resolved-scale heating in \ac{CAM}4 is applied as a hard adjustment which removes all supersaturation in a single time step, while \ac{CAM}4 deep convection has a fixed timescale (30 min) for \ac{CAPE} removal. As the time step decreases the fixed time-scale process does less to remove CAPE, while the hard adjustment does more. Since parameterized and resolved-scale deep convection have very different effects in \ac{CAM}4, changing the model time step alters the ability of these processes to compete for convective instability and manifests as strong time step sensitivity. \citet{QJ:QJ1992} presents a simple model problem to illustrate the ramifications of this time step/time-scale interaction. The example provided results in extreme model behavior due to the interaction between the dynamics and the parameterizations. While this might be described as a time step sensitivity, it is actually a sensitivity to the ratio of parameterization time scales which changes with time step. Less drastic sensitivities have been seen by other investigators which appear to be related to the time-scale ratio issue. \citet{mishra:2011} found sensitivity to time step in the average tropical rainfall amount in \ac{CAM}3 multi-year simulations, noting it was associated with the change in partitioning between convective and large-scale precipitation. \citet{reed:2012} showed a strong sensitivity in the strength of idealized tropical cyclones in high resolution \ac{CAM}5 to time step, relating it to the accompanying change to the partitioning between convective and large-scale precipitation. In both studies the time scale of the convection was not changed and thus the ratio of time scales changed. This issue of partitioning is a typical symptom observed in models that use spatial resolutions in the gray zone of cumulus convection. More discussions on the gray zone can be found in section~\ref{sec:grey}. It is worth noting that although the examples cited above are all from models that use sequential splitting, competition for resources is also a large problem for parallel splitting because it can result in unrealistically strong removal of resources. The most egregious cases of this (e.g., negative values) are typically resolved by simply rescaling tendencies to prevent over-consumption. This approach may leave more subtle cases untreated and, where applied, results in solving a different set of equations than originally intended.
Another example of the partitioning problem was shown in the work of \citet{gmd-6-861-2013}, in which case the sulfuric acid condensation and aerosol nucleation acted as two sink processes in the sulfuric acid gas budget in the \ac{ECHAM-HAM} model. The authors of that paper argued that more accurate simulations of the process rates, and consequently, more accurate near-surface concentrations of aerosol particles and cloud condensation nuclei, can be obtained when a solver handles the competing processes simultaneously without splitting.

A second scenario causing coupling problems is when one process is a source for something the other process consumes. If these processes are coupled by sequential-update splitting, the first process might push the quantity of interest to unreasonably high levels while the second process might pull it to unreasonably low levels. With parallel splitting the consuming process does not see a state immediately influenced by the source process until the following time step by which time the excess may have been modified by some other process. An example of such a push/pull problem with sequential-update splitting in \ac{CAM}5 was presented in \citet{gettelman:2015}, who note that macrophysics (condensation/evaporation + cloud fraction) is the main source of cloud water which is subsequently depleted by microphysical processes. By sub-stepping macro- and microphysics together they were able to obtain more realistic model behavior. \citet{gmd-6-861-2013} describes another push/pull problem related to sulfuric acid gas budget in \ac{ECHAM-HAM}. The study compared multiple time stepping schemes for the coupling of sulfuric acid gas production (source) and condensation (sink). Results show that when the discrete time step is long compared to the characteristic condensation time scale, sequential splitting between production and condensation leads to a substantial overestimate of the condensation rate even when the individual processes are represented with accurate solutions of the split equations. It is argued that when practical to do so, the strongly interacting sources and sinks should be solved simultaneously. A third example is presented in \citet{Beljaarsetal} for \ac{IFS}. The near-surface wind speed is mainly affected by the pressure gradient force, Coriolis force, and the turbulent friction. Sensitivity tests showed that if the turbulent diffusion coefficients are computed after the model state variables have been updated by the dynamics-induced tendencies, positive biases in the intermediate wind speeds will lead to overestimation of turbulent friction thus negative bias in the 24 hour wind forecast.

Process coupling issues can lead to large time stepping errors and strong dependence on process ordering when splitting allows processes to operate in isolation for too long. This is not uncommon in operational models where the time step is often chosen to minimize computational cost without explicitly considering accuracy. An example is given by \citet{gettelman:2015}, who note that sequential-update splitting with forward-Euler time stepping in \ac{CAM}5 microphysics creates negative cloud water when computed tendencies are multiplied by inappropriately long time steps. Another example was provided in \citet{williamson:2003}, who found that aqua-planet simulations conducted with the \ac{NCAR} \ac{CCM3} model had a single narrow peak of zonal mean precipitation at the equator when the Eulerian dynamical core was used, while simulations using the semi-Lagrangian dynamical core had a double-\ac{ITCZ} (i.e. a precipitation minimum at the equator, and two maxima straddling the equator). This sensitivity was attributed to the different time step sizes used for the physics parameterizations in the two model configurations (20 min for Eulerian, 60 min for semi-Lagrangian) rather than the dynamical cores themselves. The explanation the authors provided was that with sequential splitting, longer time steps lead to the accumulation of more \ac{CAPE}, allowing convection to initiate further from the equator. The resulting condensational heating and secondary circulation further reinforce convection away from the equator. Similar changes to \ac{ITCZ} shape in aqua-planet simulations with the \ac{CAM}3 model have also been reported by \citet{li:2011}. 

\subsection{Addressing the splitting problem}

Tighter coupling between processes is necessary to alleviate the time stepping problems noted in section~\ref{sec:dt_intro}. A crude way to do this is by simply using shorter time steps, perhaps by sub-stepping clusters of tightly-coupled processes \citep{gettelman:2015}. Sequential-tendency splitting can also be used to allow faster processes to better react to the effects of slower processes. Passing specific information from one process to another can also be useful. For example, entrainment at the top of the cloudy boundary layer in the turbulence schemes by \citet{Lock:00} and \citet{bretherton:09} is strongly affected by thermal instability diagnosed directly from radiative heating profiles. A benefit was also demonstrated in the aforementioned work on including dynamics information in the computation of turbulent surface drag \citep{Beljaarsetal}. More recently, several parameterizations have been developed which handle multiple atmospheric processes in a unified way. Two such schemes which combine turbulence and shallow convection calculations are the \ac{EDMF} approach of \citet{Siebesma:07} and the  \ac{CLUBB} approach of \citet{golaz:02a}. A scheme that unifies shallow and deep convection has been developed by \citet{Park:14}.

\section{Time step convergence}
\label{sec:time}
Due to constraints on computational resources, global weather and climate models typically use time step sizes in the range from a few minutes to an hour. Ideally, the time stepping and process-splitting methods should have sufficient accuracy at these step sizes to make the corresponding numerical errors small compared to the uncertainties associated with physically-based simplifications in the model equation system. One possible way to determine whether the time stepping accuracy is adequate is to check if the change in numerical solution caused by varying time step size stays below a practical tolerance defined through physical reasoning. Such an exercise can be interpreted as an assessment of time step convergence. In this section, convergence in the time dimension is discussed under the assumption of unchanged spatial resolution and model formulation. In this case, the asymptote of the discrete solutions -- if they converge at all -- is unlikely the best possible approximation of the real world, because inaccuracies associated with the analytic simplifications in the parameterizations and errors resulting from the spatial discretization are not alleviated by simply reducing the model time step. In other words, convergence in time alone describes the behavior of the numerical solutions of an analytically simplified, semi-discrete equation set, thus addressing only one aspect of the modeling problem.

In numerical analysis, convergence refers to the property of a numerical method that the discrete solution approaches the exact solution as the step size approaches zero. A scheme can be further characterized by its order of accuracy which describes an analytic relationship between the time step size and the local truncation error. While convergence tests (in this mathematical sense) are a standard part of dynamical core development, they have rarely been performed with model configurations combining both fluid dynamics and parameterized physics. Performing convergence tests in full-complexity models is not straightforward, for two reasons: 
\begin{enumerate}
\item In the absence of analytic solutions, a ``proxy ground truth'' is needed in a convergence analysis. Conventional convergence studies in computational fluid dynamics involve the grids for all (spatial and temporal) coordinates going to zero; applying the same test strategy to weather and climate models can cause great difficulties in the interpretation of the results, because the parameterization schemes are likely to have undesirable sensitivities to spatial resolution (which is an issue somewhat separate from time stepping error) especially when the gray zone (section~\ref{sec:grey}) is approached. Recent studies of \citet{teixeira:2007} and \citet{JAME:JAME20146} obtained reference solutions by running their models (\ac{NOGAPS} and \ac{CAM}5, respectively) with very small time step sizes. \citet{JAME:JAME20146} argued that ``convergence toward this proxy is a necessary but insufficient condition for the convergence toward the true solution''. A caveat is that the model variables might converge to an unintended and/or unphysical state when time step alone goes to zero. Additionally, physical parameterizations are often designed to work within a particular range of time steps and using them outside of that range may violate physical assumptions. For example, most climate models assume that supersaturation with respect to liquid water is removed instantaneously, which is not true on timescales less than about $10$~sec (cf \citet{squires:1952}). Thus it is probably better to interpret the difference with respect to the reference solution as a metric of time step sensitivity rather than as error relative to the true solution of the chosen equation system.
\item Another difficulty in time step convergence analysis is that the expected behavior of operational models is not yet well established. Note that the concept of ``order of accuracy'' was originally developed for deterministic differential equations under the assumption that the solution is time-differentiable at least up to a certain order, but in full-fledged weather and climate models the condition is not always fulfilled. \citet{hodyss:2013} used numerical simulations of the diffusion-advection equation to demonstrate that when the time stepping scheme does not resolve the parameterized physical processes, the numerical solutions will behave as predicted by stochastic theory, resulting in a substantially reduced convergence rate. The implication of their results is profound: since weather and climate models include impactful processes (e.g., microphysics, turbulence) with time scales of seconds or smaller while models are typically run with time steps several minutes in length, the originally expected first- (or higher-) order convergence might not be realizable. In deterministic numerical analysis, convergence is defined in the limit of small time steps, when model response to time step change can be approximated by a Taylor series truncated at the order of convergence. The large time steps used operationally in weather and climate models may fall outside the domain of validity of the truncated Taylor approximation, so the practical impact of reducing the model time step may be very different than predicted by deterministic numerical analysis. The slow convergence of the CAM5 simulations described by \citet{JAME:JAME20146} is likely an indication that the default model time step is far too long to resolve all the intrinsic time scales associated with the equation system. How to assess and improve solution accuracy in such a situation is a topic that requires further investigation
\end{enumerate}

In addition, the traditional truncation error analysis often quantifies the numerical error of a discretization scheme in a single time step, i.e. the local truncation error, while in practice the global error accumulated in all the steps leading to a fixed simulation time is perhaps a more relevant metric. \citet{teixeira:2007} conducted a number of simulations with different time step sizes using \ac{NOGAPS}, a  \ac{QG} model, and the Lorenz equations. They found that \ac{NOGAPS} converged at first order near the start of their simulations, but the chaotic nature of nonlinear dynamical systems eventually caused simulations with different time steps to diverge into uncorrelated sequences of weather events. When one attempts to examine convergence rates beyond the first few steps of a simulation, uncertainties associated with the nonlinear nature of the equation system (``internal variability'') need to be taken into account.

In operational weather and climate models, the magnitude of error obtained at a given resolution or given cost is the most important and practical measure of the quality of the time stepping method. Nevertheless, despite the abovementioned complication, a convergence analysis might still provide useful information about the numerical properties of the discrete model system, especially when the results deviate from the expected behavior. For example, \citet{JAME:JAME20146} found that in \ac{CAM}5 the parameterizations that converge slower also have stronger time step sensitivity. In their case, the convergence rate provides a clear hint on which components of the model have inadequate numerical treatment thus require more attention in future development. 
Using the single-column version of a model to isolate the impact of physics parameterizations from fluid dynamics problems may help identify whether time stepping problems are related to physics, dynamics, or the interaction between the two, however the single-column approach requires a large number of cases to capture the variety of physical processes approximated in the models. Sub-stepping an individual process or a cluster of processes is a widely used strategy for improving stability and accuracy for faster components in a system. For example, the spectral-dynamical core in CAM5 uses multiple levels of sub-stepping for the adiabatic fluid dynamics, resolved-scale tracer transport, and numerical diffusion \citep[cf. Table 1 in][]{JAME:JAME20146}. \citet{gettelman:2015} noted that using smaller time steps for the stratiform cloud parameterization had a positive impact on the model behavior. From the perspective of convergence analysis, sub-stepping clusters of problematic processes while keeping all aspects of other processes untouched can be useful for finding problems with certain schemes or process coupling. Replacing the time-integration method for certain schemes may play a similar role to sub-stepping. However, it should be kept in mind that while sub-stepping can improve the stability and accuracy of individual processes, it cannot address the splitting problem discussed in section~\ref{sec:arti} unless the strongly interacting processes are sub-stepped as a cluster.

As mentioned above, time step convergence tests of full-complexity models is a rarely-conducted exercise in the weather and climate modeling community. It will be interesting to see the outcome of the ongoing efforts in this direction.

With these real world issues and examples in mind the paper now proceeds into a more theoretical area, a mathematical analysis approach to the coupling.

\section{ Insights from models with simplified equation sets }
\label{sec:simplified_eq}

Two examples of \ac{PDC} are discussed below  where the resolved scale behavior is strongly dependent on the subgridscale dynamics. This analysis highlights situations where the combination of resolved and subgrid terms is critical, e.g. in representing the total transport as the sum of resolved and subgrid transport. As the averaging scales are reduced, the subgrid contribution will reduce and be taken over by the resolved contribution. 
\subsection{Interaction of convection with balanced dynamics}
\label{sec:simplified_eq_1}
In this case the spatial averaging scale is relatively large, and so the semi-geostrophic model, which is an accurate approximation to the governing equations on large scales, \cite{Cull06}, can be used as a proxy for the evolution of the spatially averaged equations. The behavior of this model can then be compared with solutions of the true governing equations with a much finer averaging scale which resolves convection explicitly. This then has implications for the design of models with parameterized convection.

The semi-geostrophic model includes
the effect of large static stability variations, which are essential in
considering interactions with convection. For illustration the in-compressible Boussinesq form of the equations in Cartesian geometry is used. Following \cite{QJ:QJ200312958915}, the equation for the ageostrophic wind is written as
\begin{equation}
\label{seq1}
{\bf Q}{\bf u}_{ag}+\frac{\partial}{\partial t}\nabla p={\bf H},
\end{equation}
where
\begin{eqnarray}
\label{seq2}
{\bf Q}=\left(\begin{array}{ccc}fv_{gx}+f^2&fv_{gy}&fv_{gz}\\
-fu_{gx}&f^2-fu_{gy}&-fu_{gz}\\
g\theta_x/\theta_0&g\theta_y/\theta_0&g\theta_z/\theta_0\end{array}\right),\\
{\bf H}=\left(\begin{array}{c}-f{\bf u}_g\cdot\nabla v_g+F_1\\f{\bf u}_g\cdot\nabla
u_g+F_2\\-g{\bf u}_g\cdot\nabla\theta/\theta_0+S\end{array}\right).\nonumber
\end{eqnarray}
Here ${\bf u}=(u,v,w)$ is the velocity, with suffix $g$ indicating geostrophic
and suffix $ag$ indicating ageostrophic values. Suffices $x,y$ and $z$ indicate spatial derivatives. $f$ is the Coriolis parameter,
$g$ is the acceleration due to gravity and $\theta$ is the potential temperature with
reference value $\theta_0$. $F_1,F_2$ and $S$ are momentum and thermodynamic forcing terms respectively.

Under semi-geostrophic dynamics, the ageostrophic flow is determined diagnostically, and thus represents a response to the dynamical and physical forcing represented in equation (\ref{seq1}). The strength of the response is determined by the eigenvalues of ${\bf Q}$, which represent the inertial and static stability of the atmospheric state. The geostrophic state would be expected to be described by the resolved flow in numerical models. However, the ageostrophic circulation required to maintain geostrophic balance would include subgrid-scale transports as well as resolved ageostrophic transport.

In the presence of moisture, the static stability is reduced by latent heating. This could be expressed, neglecting precipitation, by replacing $\theta$ by the equivalent potential temperature $\theta_e$ in saturated regions. In the presence of moist instability, {\bf Q} would then have a negative eigenvalue. As illustrated in \cite{holt}, this will result in convective transport rather than smooth vertical motion. The effect is that convective mass transport would replace the ascending branch of the ageostrophic circulation, while the downward ageostrophic circulation would be a smooth transport. 

This prediction is illustrated using a convection-permitting simulation performed as part of the \ac{EMBRACE} %
\texttt{http://cordis.europa.eu/project/rcn/99891\_en.html}
 project. The simulation uses a configuration similar to that used operationally at the Met Office for \ac{UK}-area short-range weather prediction \citep[see][for details]{Holloway:2012} but with changes made to improve the representation of tropical convection and gravity waves. It has a horizontal resolution of $2.2$~km with a large $8800$~km by $5700$~km domain centered on the tropical Indian ocean and $118$ vertical levels with a $78$~km lid. Within its domain the convection-permitting simulation was run freely after being initialized from the operational Met Office global model analysis valid at 0000~\ac{UTC} on 18~August~2011. The lateral boundary conditions were provided every time step by a global model that was reinitialized from Met Office operational analyses every 6~hours. The data presented here was taken from 0000~\ac{UTC} on 30~August~2011 and hence the convection permitting simulation was fully spun up.

The high resolution gridpoints are classified as cloudy or dry depending on the presence or not of cloud condensate: the cloudy areas are further subdivided into ascending and descending. The high resolution gridpoints are then aggregated onto a 24~km grid (a typical resolution at which convective parameterization is used) so that for each 24~km gridpoint a cloudy and dry mass flux is obtained, cloudy updrafts and downdrafts, and also the total large-scale mass flux.

\begin{figure*}
  \centering{\includegraphics{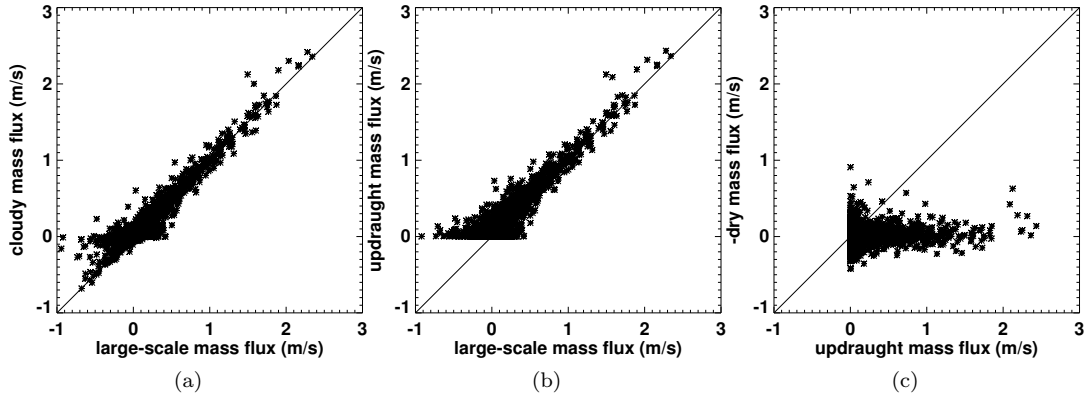}
}
\caption{Scatter plots of cloudy mass flux against large-scale mass flux (a), cloudy updraft mass flux against large-scale mass flux (b), and  minus dry mass flux against cloudy updraft mass flux (c). All in units of m s$ ^{-1}$. The data is taken from a height of $3195$~m and is averaged in the horizontal to scale of $24$~km. Met Office Unified Model. }
  \label{fig:seq1}
\end{figure*}

Figure~\ref{fig:seq1} shows that, for $24$~km gridpoints that have some cloud, there is a close match between the total large-scale mass flux and the cloudy mass flux and hence most of the vertical motion happens within the cloudy areas (cf. section~\ref{sec:modlelimgrey}). The values of the dry mass flux are unrelated to the cloudy updraft mass flux. This means that there is no local compensating subsidence within the $24$~km gridbox to match the cloudy updraft mass flux as is usually assumed in convective parameterization. The subsidence is instead spread over the whole domain. This is in agreement with the idea that the large-scale ascent is represented by convective plumes, while the subsidence is spread over a much wider region. This suggests that a radical rethink of (convective) parameterization strategy is required.

\subsection{Interaction of the boundary layer with balanced dynamics}
\label{sec:simplified_eq_2}
In this case the characterization of a simple model as the asymptotic limit of the full equations is exploited, and used to compare the effectiveness of different coupling strategies. A large scale balance is defined, which should be represented in the resolved numerical solutions, while the circulation required to maintain it will be described by both resolved and subgrid-scale transports. The inclusion of the boundary layer makes a fundamental change to the  large scale balance because of the need to satisfy the no-slip boundary condition. Thus the balance is defined by the Ekman relations
\begin{eqnarray}
\label{seq3}
\frac{\partial p}{\partial x}-fv_e=F_1({\bf u}_e),\\
\frac{\partial p}{\partial y}+fu_e=F_2({\bf u}_e).\nonumber
\end{eqnarray}
$(u_e,v_e)$ are the components of the Ekman velocity, and $F_1$ and $F_2$ represent the parameterized friction terms, which will depend on the horizontal momentum as indicated, as well as the thermodynamic structure. These equations can be solved for ${\bf u}_e$ given that ${\bf u}_e={\bf u}_g$ at the top of the boundary layer and is zero at the ground.
 
\cite{beare13} derive equations analogous to equation~\ref{seq1} for the circulation required to maintain Ekman balance in time in the presence of dynamical and physical forcing. The ageostrophic circulation in semi-geostrophic theory is a second order accurate approximation in Rossby number to the velocity in the Euler equations. However, the equivalent circulation in the boundary layer is only first order accurate, as is the Ekman balance itself.
 
Now the effectiveness of schemes to couple the boundary layer with the balanced dynamics is demonstrated by following the method of \cite{cull07}. This experiment is described in detail by \cite{beare15}.  A vertical slice model is used to construct a sequence of solutions of the boundary layer driven by a baroclinic wave where the Rossby number $U/fL$, with $U$ and $L$ denoting horizontal velocity and length scales, respectively, is progressively reduced. This is achieved by maintaining the same initial structure in the pressure and potential temperature while simultaneously increasing the Coriolis parameter and reducing the wind speed. The difference between the predicted circulation and the solution of the hydrostatic equations is then calculated. The expected result is second order convergence outside the boundary layer and first order inside. However, the boundary layer is found to become shallower as the  \ac{Ro} is reduced, giving an overall convergence rate of \ac{Ro}$^{1.7}$. 

The results are compared using three numerical implementations. The control simulation uses a standard implicit time stepping, but the mixing coefficients $F_1$ and $F_2$ are evaluated only at the beginning of the time step. The \cite{wood07} scheme is a stable single step scheme which is unconditionally stable and second-order accurate. This is achieved by assuming a polynomial dependence of $F_1,F_2$ on wind speed. The $K$-update scheme includes the updated value of the boundary layer mixing coefficient at the new time level in each time step as described by \cite{QJ:QJ200312958915} as well as the more accurate representation of the diffusion process in \cite{wood07}. This allows it to represent the balanced solution more accurately.

\begin{figure*}
  \centering{\includegraphics[scale=0.5]{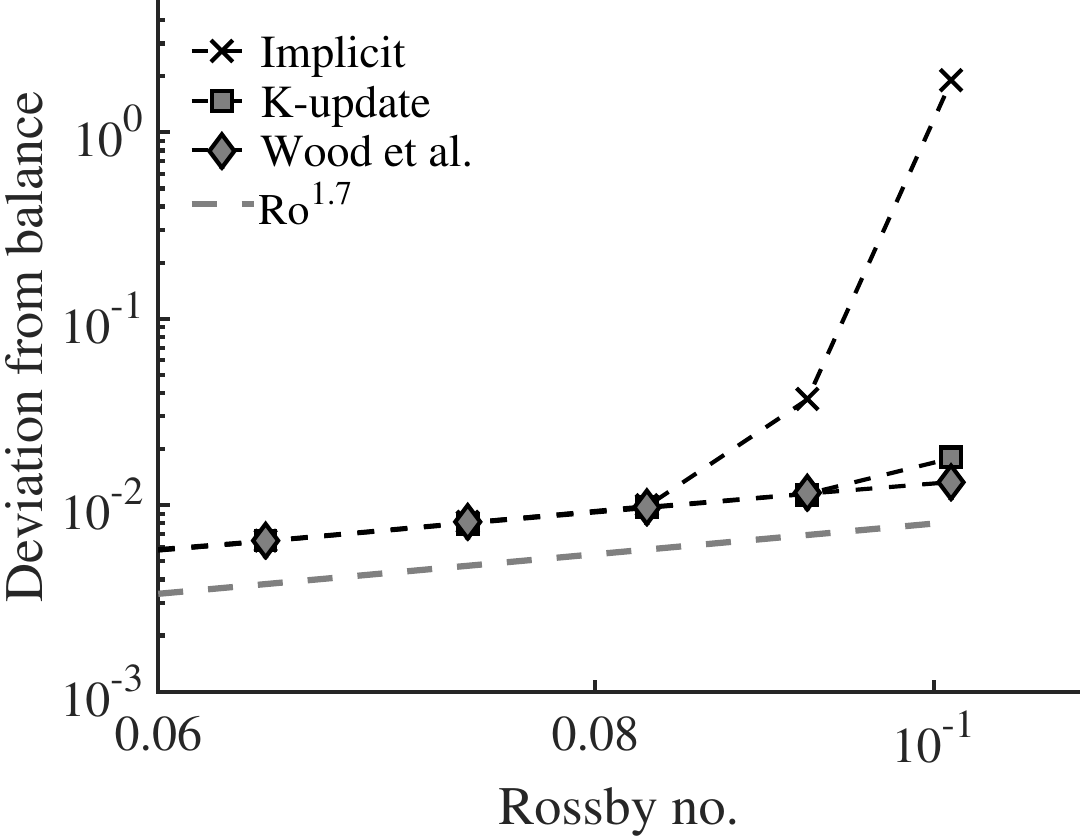}
  }
  \caption{Convergence to balanced solution of the vertical slice primitive equation simulations \citep{beare15} for different  time stepping schemes: Implicit, K-update and \cite{wood07}. Ro$^{1.7}$ is shown in gray for reference.
.}
  \label{fig:seq2}
\end{figure*}

Figure~\ref{fig:seq2} shows the difference between primitive equation simulations using different boundary-layer time stepping schemes and the balanced model. At smaller Rossby numbers, all primitive equation models follow the ideal Ro$^{1.7}$ line. However, above Ro~=~$0.08$, the primitive equation model using the  implicit scheme starts to deviate significantly above the ideal line, and no longer converges at the required rate. The primitive equation model using the K-update scheme deviates slightly above the ideal line at Ro~=~$0.1$. The \ac{HPE} model using the \cite{wood07} scheme follows the ideal 
Ro$^{1.7}$ line for the range of Ro shown. Both the K-update and 
\cite{wood07} schemes account for the variation of the boundary-layer diffusion across the time step, giving the improved convergence properties compared to the Implicit scheme. The deviation from the Ekman-balanced models thus exposes differences in the numerical methods employed. 

\subsection{Summary}
Two approaches of validating methods of dynamics-physics coupling were  demonstrated, given that the required averaged solution of the full equations cannot be described exactly as the solution of a set of partial differential equations. Section~\ref{sec:simplified_eq_1} illustrated that the validation of subgrid models should be against the averaged data from much higher resolution models. How the suggested changes translate into the full model also needs to be explored. Section~\ref{sec:simplified_eq_2} showed how methods of coupling of subgrid models can be validated by the accurate reproduction of asymptotic limits where subgrid transports are a key part of the limit solution. A future aim is to create protocols and setups for more complex models.

\section{Analyzing the coupling of dynamical cores with a hierarchy of GCM test cases}

\label{sec:simple}
One of the recurring questions is: Which \ac{PDC} strategy is better? The answer depends crucially on the objective of the model run. Is it a climate run or a weather forecast? Is the model already severely time step restricted, such as Eulerian formulations, or are long time steps permitted, as in semi-implicit semi-Lagrangian models? The former may be less susceptible to coupling errors, assuming the physics and dynamics time steps are not too disparate, due to the higher temporal resolution and less scope for non-linear evolution (or splitting error). But even when these questions have been answered, in the full model context it is far from trivial to say which is better since the large number of factors involved quickly blur the answers. Therefore, testing is essential, and  it is proposed that a hierarchy of idealized \ac{GCM} test cases gives easier access to an improved understanding of the coupling mechanisms.

\subsection{Idealized testing of GCMs}
Full model testing has been discussed above and, for example, \cite{JAME:JAME20146} proposed various analysis techniques to better understand the impact of the physics time step on the model behavior. In an idealized framework, however, the parameterizations and lower boundary conditions are more constrained, which  exposes the impact of the physics-dynamics coupling strategy on the simulation in a clearer way.

\subsubsection{The different nature and sources of error}

If models did not require tuning~\citep{doi:10.1175/BAMS-D-15-00135.1}, the answers would perhaps be more obvious. However, if a novel coupling method is implemented in an already tuned model, the solution is likely to be worse for the new coupling method if the model is then not re-tuned, even if the new coupling strategy would lead to a superior solution in the absence of tuning. Model tuning inevitably tunes against errors that are independent of the parameters tweaked in the tuning process (i.e., compensating errors). In this case, multiple errors may exist, but the superposition of errors introduced to minimize other errors may result in ``shadowing of errors'' if only the final solution is taken into account during tuning processes. Remove one of these errors and the result will be worse, despite having eliminated an error. For example, removing (or reducing) errors in the coupling of a mature model may result in a degraded final solution for these reasons. This highlights a key challenge in \ac{PDC}: Not one single experiment will yield all the answers. The different techniques presented here have to be taken as a cohort of interrogation. Each has to be interpreted under there individual limitations. Combined it should be possible to derive clearer guidelines and understanding of the complex interactions. Not one analysis method is valid or one limitation invalidates the other. In a slightly modified version of Abraham Kaplans  ``The Conduct of Inquiry'': The models are undeniably beautiful, and a man may justly be proud to be seen in their company. But they \textbf{have} their hidden vices. The question is, after all, not only whether they are good to look at, but whether we are able to interpret their results in the context of their limitations.

\subsubsection{The \ac{GCM} test case hierarchy}

Due to the interconnected sources of error illustrated above it seems reasonable to implement and standardize an idealized testing protocol. It should be idealized in such way that the complexity of physical parameterizations is present in the forcing, but not in the implementation and that the implementation is generalized, allowing for direct comparisons between models.
For the dynamical core, tests with idealized forcing exist, such as the Held-Suarez test case \citep{suHe1}. The Held-Suarez forcing was formulated for a dry and flat planet and includes a thermal relaxation mechanism and low-level Rayleigh friction. These mimic the effects of radiation and boundary layer mixing, respectively. However, the adjustment processes in the Held-Suarez test case are rather slow and do not challenge the physics-dynamics coupling sufficiently. A missing key ingredient is moisture. The latent heat exchanges due to water phase transitions are desirable in order to challenge the coupling mechanisms. 

The ``simple-physics'' package by \cite{JAME:JAME57} makes progress in this aspect. It incorporates bulk aerodynamic surface fluxes and diffusive boundary layer mixing processes of heat, moisture and momentum, a large-scale condensation scheme without a cloud phase, and utilizes an ocean-covered surface with prescribed  \acp{SST} as a lower-boundary condition. The Fortran source code is publicly available (\texttt{https://earthsystemcog.org/projects/dcmip-2012/}, click on 'Fortran Routines' in left navigation bar and download the attached 'simple physics suite' at the bottom of the page), removing the uncertainty of the implementation, and the suite is simplistic enough to be easily reproduced within varying model frameworks. However, the simple-physics package lacks radiation and is therefore only suitable for short-term simulations. 
This was remedied by \cite{thatcher:16} who combined the ideas of the \cite{JAME:JAME57} simple-physics package and the Held-Suarez forcing to create a moist version of the Held-Suarez test. The resulting \ac{MITC} with Newtonian thermal relaxation mimicking ``radiation'' is suitable for long-term simulations and has been shown to reveal the intricacies of the physics-dynamics coupling as further highlighted in section~\ref{subsec:mitc}. \ac{MITC} can be considered a moist idealized test of intermediate complexity. The MITC Fortran routine is available as a supplement to the journal article from \texttt{http://www.geosci-model-dev.net/9/1263/2016/}. 

The next step in the test case hierarchy points to simplified physics formulations with a radiation scheme and unconstrained SSTs that are e.g. determined by a slab ocean model (also called ``mixed-layer'' model).  \cite{frierson06} presented a gray-radiation \ac{GCM}, which possesses desirable ingredients such as radiation, an interactive slab ocean, large-scale precipitation, and surface/boundary layer schemes. However, the physics suite is not sufficiently documented to be easily reproducible and comparable to other models. 
If more realistic ocean temperatures are desired, a slab ocean scheme can also be augmented with a set of specified surface flux adjustments (commonly called ``q-flux adjustments''). These can be added to the slab model's temperature tendency equation at each time step in order to maintain a seasonal cycle of realistic ocean temperatures.

A final step in the idealized model hierarchy are long-term ``aqua-planet'' simulations on a flat and ocean-covered Earth that utilize the complex physical parameterization package of a \ac{GCM}. The lower boundary condition can either be based on prescribed \acp{SST} as in \cite{NH2000ASL} or a slab ocean approach with predicted \acp{SST} as in \cite{lee:08}. Aqua-planet simulations are popular for idealized climate studies. Here, we demonstrate that they can also provide insight into the delicate interplay between the physical parameterizations and the numerical schemes of dynamical cores with their associated diffusion (section~\ref{subsec:aqua}).
Ideally, in between the two well observed and understood boundary conditions, \ac{SST} and incoming shortwave radiation at the top of the atmosphere,
as much as possible should be left to the model, ie. variables should be allowed to propagate freely and not be prescribed or constrained to reference profiles or background states. If a certain coupling scheme or model formulation has an impact on the Hadley circulation, for example, then the test case should be able to show a trend towards this.

\subsection{Simplified physics assessments}
\label{subsec:mitc}
\begin{figure}
 \noindent\includegraphics[width=20pc]{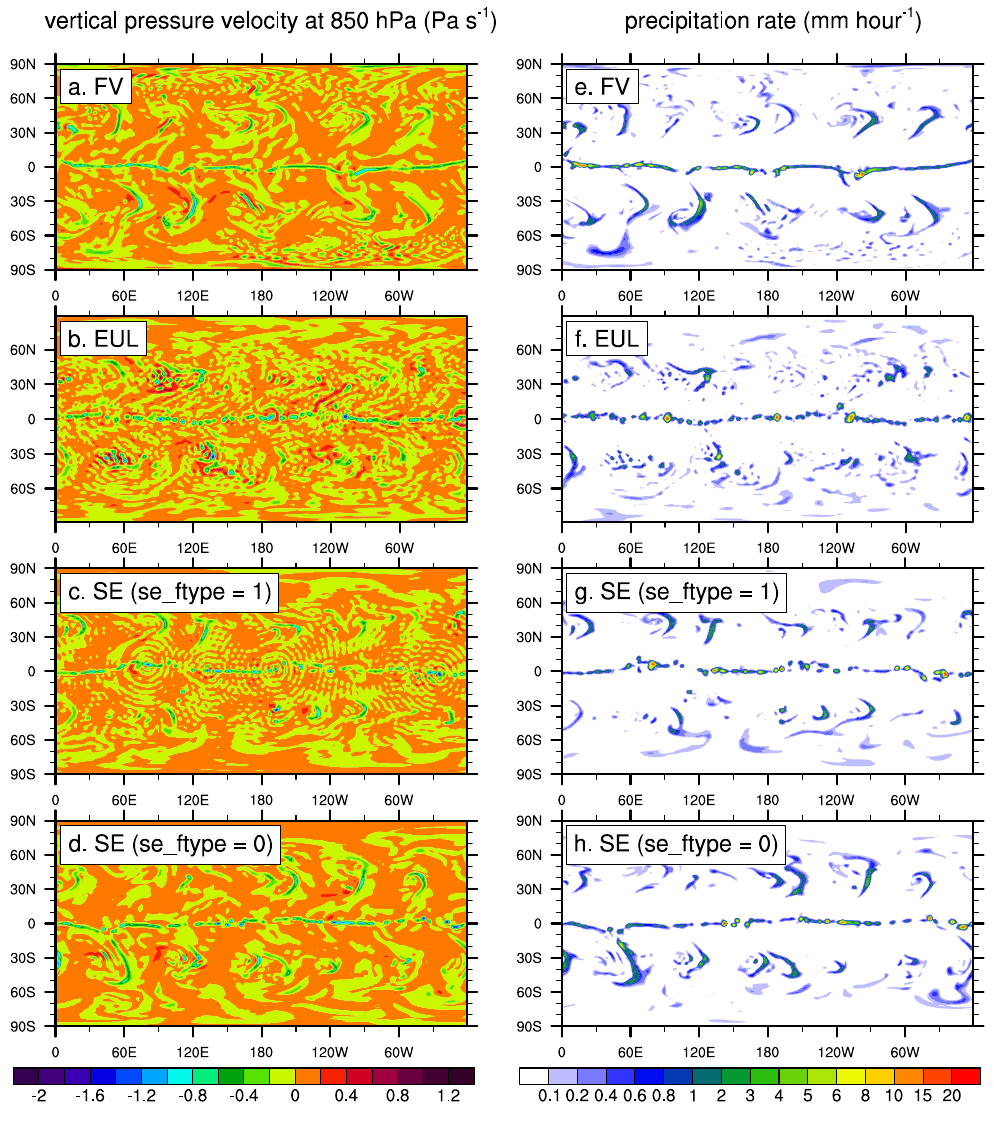}
  \vspace{-0.6cm}
 \caption{Snapshots of instantaneous (left) 850 hPa vertical pressure velocities and (right) precipitation rates in \ac{MITC} simulations with
 the (a,e) \ac{CAM}-\ac{FV}, (b,f) \ac{CAM}-EUL and (c,d,g,h) \ac{CAM}-\ac{SE} dynamical cores. se\_ftype = 1 (c,g) denotes a physics-dynamics coupling with the long physics time step, se\_ftype = 0 (d,h) couples with a sub-cycled, short dynamics time step.
The physics time steps are $1800$~s (FV, SE) and $600$~s (EUL), the dynamics time steps are 180~s (FV), 600~s (EUL) and 300~s (SE). In the case of \ac{SE} with se\_ftype=0 the forcing was gradually applied every $300$~s. The EUL dynamical core is coupled to the physics in a process-split (parallel) way, the \ac{SE} and \ac{FV} physics-dynamics coupling is time-split.}
  \label{fig:mhs}
 \end{figure}

Figure~\ref{fig:mhs} displays an example of how the \ac{MITC} approach by \cite{thatcher:16} can provide information about the physics-dynamics coupling strategy.
The figure shows instantaneous, randomly selected snapshots of the 850 hPa vertical pressure velocities and precipitation rates in \ac{MITC} simulations with the \ac{CAM}5 model \citep{CAM5}. The depicted \ac{CAM}5 dynamical cores are the \ac{FV} model \citep{lin:04}, the spectral transform Eulerian (EUL) dynamical core, and the \ac{SE} model \citep{TF2010JCP,dennis:12}. These are run at the horizontal resolutions $1^\circ \times 1^\circ$ (\ac{FV}, $\approx$111~km), the triangular truncation T85 with a quadratic Gaussian grid (EUL, $\approx$156~km), and in the ``$ne30np4$'' (SE) configuration which corresponds to a grid spacing of about 111~km. All dynamical cores use the same 30 vertical levels. Their positions are documented in the Appendix of \cite{JAME:JAME57}. 

The three dynamical cores are coupled to the identical MITC physics package \citep{thatcher:16} and run for multiple years. Within the MITC physics package, the coupling strategy of the various physical processes follows the sequential-update approach which is also detailed in \cite{thatcher:16}. However, the physics-dynamics coupling strategies differ. The \ac{FV} dynamical core (Figures~\ref{fig:mhs}a,e) with a dynamics time step of $180$~s  is coupled to the physics package in a time-split (sequential) way and applies the physical forcings every $1800$~s (physics time step). The EUL dynamical core (Figures~\ref{fig:mhs}b,f) is coupled to the physics in a process-split (parallel) way. EUL applies the physical forcings every $600$~s which is identical to EUL's dynamics time step. The \ac{SE} dynamical core (Figures~\ref{fig:mhs}c,d,g,h) with a dynamics time step of $300$~s is coupled to the physical parameterizations in a time-split way with a physics time step of $1800$~s as \ac{FV}. However, two coupling options exist in \ac{SE} which either apply the physical forcings as a sudden adjustment after the long  $1800$~s physics time step (se\_ftype = 1) or gradually within the sub-cycled dynamical core (se\_ftype = 0) every $300$~s.

Figures~\ref{fig:mhs}(c,d,g,h) document that the choice of the coupling strategy in \ac{CAM}5-\ac{SE} has significant impact on the simulation. The intense gridscale (or gridpoint) storms \citep{QJ:QJ1992}, that develop along the equator in all models (seen in the precipitation rates in the right column), lead to a circular gravity wave ringing patterns in the 850 hPa vertical pressure velocity $\omega$ in \ac{CAM}5-\ac{SE} when coupled with the long $1800$~s physics time step (se\_ftype = 1, Figure~\ref{fig:mhs}c). The centers of the circular $\omega$ patterns coincide with the positions of the strongest precipitation rates in Figure~\ref{fig:mhs}g, which suggests that the intense latent heat release at these locations initiates the gravity wave noise. The gravity wave response to the impulsive physical forcing is large-scale, so that the explicitly-applied diffusion in \ac{CAM}5-\ac{SE} does not filter out its propagation. \cite{thatcher:16} found that the gravity wave noise can be remedied when changing the coupling strategy in \ac{CAM}-SE. In case of se\_ftype = 0 (Figures~\ref{fig:mhs}d,h) the physical forcing tendencies are gradually applied within the \ac{CAM}-SE dynamical core every $300$~s. The strong grid-scale storms are still present in the precipitation field (Figure~\ref{fig:mhs}h). However, the more gradual forcing reduces the latent heat impulses and leads to a smooth vertical pressure velocity (Figure~\ref{fig:mhs}d). Similar sensitivities to the  se\_ftype setting were also found in full-complexity CAM-SE climate simulations (Peter Lauritzen (NCAR) personal communication, 2015). Therefore, the CAM-SE se\_ftype default was switched from 1 to 0. This shows that simpler modeling frameworks help expose the causes and effects of the physics-dynamics coupling choices.

It is also informative to compare these \ac{CAM}-SE characteristics to the alternative FV and EUL dynamical cores. As SE (se\_ftype = 1), the FV model (Figures~\ref{fig:mhs}a,e) also adjusts the state variables with the long 1800~s physics time step and experiences equatorial grid-point storms of similar magnitude (Figure~\ref{fig:mhs}e). However, the damping characteristics of the two dynamical cores differ \citep{jablonowski:11} and FV can more effectively damp grid-scale noise due to its built-in, local monotonicity constraints. Therefore, FV distributes the large latent heating impulses more smoothly which leads to a smooth distribution of its vertical pressure velocity (Figure~\ref{fig:mhs}a). In contrast, the EUL model is built upon a global spectral numerical method which is known for its difficulty representing sharp contrasts locally. Here, the large latent heating impulses near the peak precipitation rates (Figure~\ref{fig:mhs}f), lead to the so-called Gibbs ringing effect (see also \cite{jablonowski:11}). The Gibbs ringing is visible in EUL's vertical pressure velocity field (Figure~\ref{fig:mhs}b) and manifests itself as a noisy pattern (broken contours). The noise is even present in the midlatitudinal regions where organized precipitation bands should dominate. EUL's shorter 600~s physics time step (in comparison to the 1800~s used in FV and SE) is not able to prevent these numerical Gibbs oscillations.

\subsection{Aqua-planet assessments}
\begin{figure}
 \noindent\includegraphics[width=20pc]{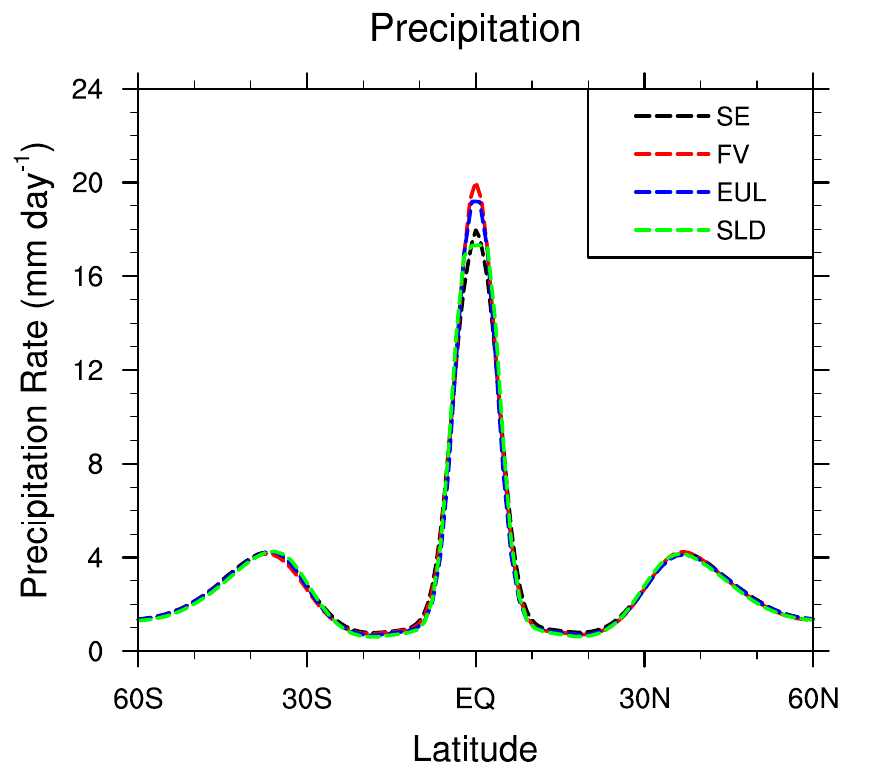}
  \vspace{-0.5cm}
 \caption{2-year-mean zonal-mean precipitation rate in four aqua-planet simulations with the \ac{CAM}5 dynamical cores \ac{SE} (111~km), \ac{FV} (111~km), EUL (T85), SLD (T85) and the default \ac{CAM}5 physics package.}
  \label{fig:aqua}
 \end{figure}

\begin{figure*}
 \noindent
 \includegraphics[angle=-90]{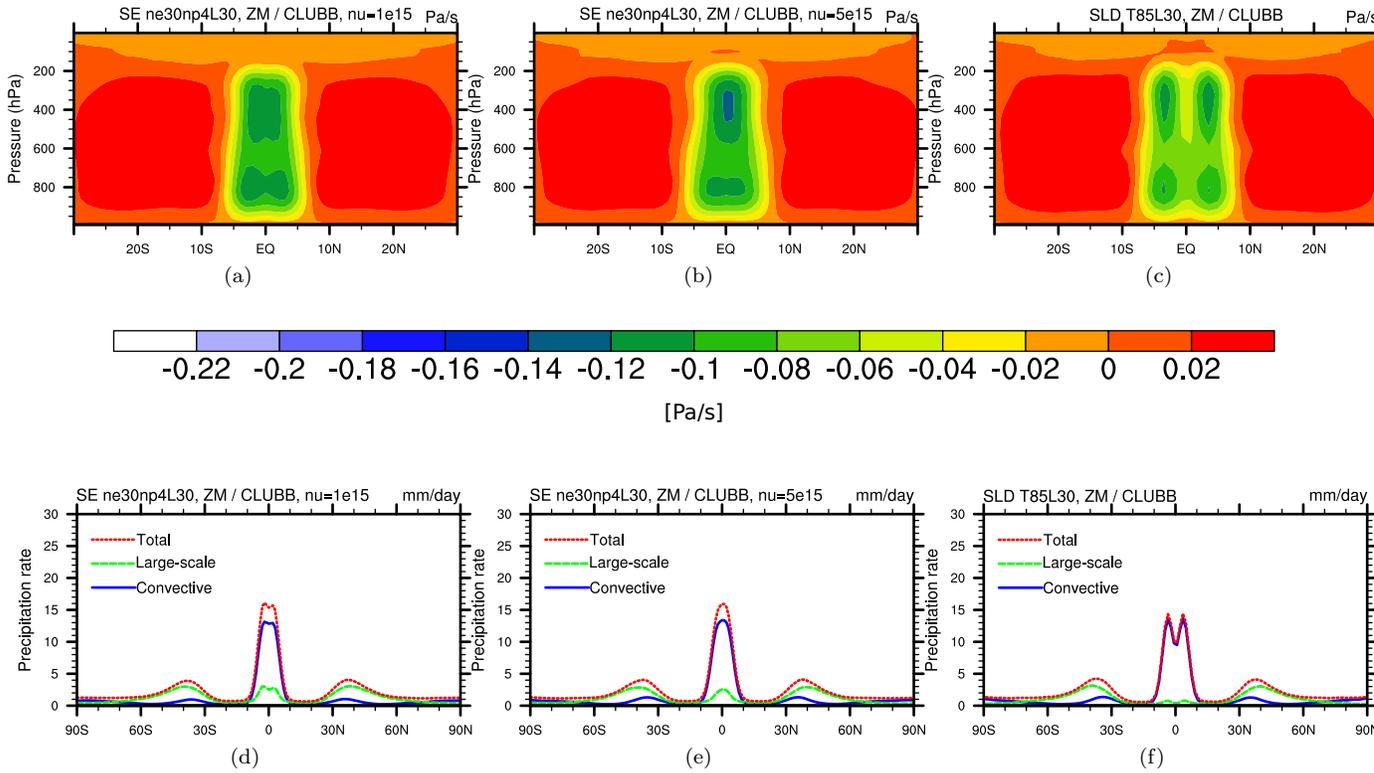}
 \caption{Aqua-planet simulations with the alternative \ac{CLUBB} \ac{PBL}, macrophysics and shallow convection scheme in \ac{CAM}5. Top: Latitude-pressure cross section of the one-year-mean zonal-mean vertical pressure velocity in the tropics for the dynamical cores (a) \ac{SE} with diffusion coefficient $1\times10^{15}$~m$^4$s$^{-1}$, (b) \ac{SE} with diffusion coefficient $5\times10^{15}$~m$^4$s$^{-1}$ and (c) SLD without explicit horizontal diffusion. Bottom: One-year-mean zonal-mean precipitation rates of the three runs, split into total (red), large-scale (green) and convective (blue) precipitation.}
  \label{fig:aqua-CLUBB}
 \end{figure*}

Another example of how full-physics aqua-planet simulations can give insight into the physics-dynamics interplay is shown in Figures~\ref{fig:aqua} and \ref{fig:aqua-CLUBB}. The figures provide information about the shape of the \ac{ITCZ} in \ac{CAM}5 aqua-planet simulations with prescribed \acp{SST} (CONTROL case in \cite{NH2000ASL}). As in section~\ref{subsec:mitc} the \ac{CAM}5 dynamical cores EUL, \ac{FV} and \ac{SE} are assessed at the resolutions T85 (EUL) and $111$~km (\ac{SE}, \ac{FV}) with 30 levels. In addition, the figures include the \ac{CAM}5 spectral transform semi-Lagrangian (SLD) T85 dynamical core. All model simulations are run for 2.5 years, and the first six months are disregarded (spin-up period). The models use the dynamics time steps 300~s (SE), 180~s (FV), 600~s (EUL) and 1800~s (SLD), which are paired with the physics time steps 1800~s (SE, FV, SLD) and 600~s (EUL). 

The shape of the \ac{ITCZ} in aqua-planet simulations has been a topic of debate for over a decade. Some models show a single equatorial peak of the \ac{ITCZ} precipitation rate whereas other models are characterized by a double \ac{ITCZ} in the subtropics. \cite{blackburn:13b} even called the double \ac{ITCZ} one of the ``modern modeling mysteries''.
The suggested mechanisms that govern the shape and strength of the \ac{ITCZ} vary widely and are ambiguous. \cite{williamson:2003} found a dependence on the physics time step, time stepping scheme, the dynamical core and the strength of the horizontal diffusion. \cite{mishra:08} discussed the ITCZ time step dependencies and physics changes, \cite{rajendran:13} discussed the \acp{SST} impact on the ITCZ, \cite{lee:03} and \cite{moebis:12} investigated the role of the convection scheme, \cite{TELA:TELA339} reported on the sensitivities to horizontal resolution, and \cite{landu:14} discussed the \ac{ITCZ} sensitivity to two dynamical cores, their resolutions and strengths of the low-level moisture transports. More recently, \cite{medeiros:15,medeiros:16} compared the ITCZs in the aqua-planet configurations of models that participated in the Coupled Model Intercomparison Project Phase 5 (CMIP5) and provided an aqua-planet reference solution for NCAR's CAM5.3 (version 5.3) model.

\label{subsec:aqua}

Figure~\ref{fig:aqua} depicts the time-mean (averaged over the last 2 years) zonal-mean precipitation rate in all four \ac{CAM}5 dynamical cores in aqua-planet mode which are driven by the identical \ac{CAM}5 physical parameterization package. The physics package is described in \cite{CAM5}. Here, we highlight that it contains the  \ac{UW} moist turbulent \ac{PBL} scheme by \cite{bretherton:09} which is based on an assessment of the turbulent kinetic energy. This \ac{PBL} scheme is tightly linked to the \ac{UW} shallow convection parameterization, developed by \cite{park:09}, through a cloud-base mass flux. In addition, the \ac{CAM}5 deep convection
scheme is based on the mass-flux parameterization by \cite{zhang:95}, which has been enhanced by \cite{richter:08} and \cite{neale:08} to account for convective momentum transport and dilute plumes.

Figure~\ref{fig:aqua} shows that the precipitation rates in all four aqua-planet simulations are remarkably similar. They all show a single \ac{ITCZ} and equatorial peaks that range between $17.5$ -- $20$~mm~day$^{-1}$. This is in sharp contrast to the assessments by \cite{blackburn:13b} who intercompared 16 different model simulations that participated in the  \ac{APE}  \citep{blackburn:13a}. The peaks in the \ac{APE} models ranged from 10-34 mm day$^{-1}$ with an almost even split between single versus double \ac{ITCZ} models. Since the \ac{APE} models are characterized by vastly different dynamical cores, resolutions, physical parameterizations and coupling strategies this makes it very difficult to distinguish between causes and effects. When taking Figure~\ref{fig:aqua} into account though, it seems feasible that most differences in \ac{APE} models are likely triggered by different physical parameterizations.

Here a single aqua-planet framework is promoted as a ``control environment'' for idealized assessments of the physics-dynamics interplay. An example is given in Figure~\ref{fig:aqua-CLUBB} which intercompares the \ac{CAM}5 \ac{SE} (111~km) and SLD T85 dynamical cores with 30 levels when coupled to the alternative physical parameterization scheme \ac{CLUBB} \citep{golaz:02a,golaz:02b,bogenschutz:12,bogenschutz:13}. \ac{CLUBB} replaces \ac{CAM}5's default \ac{PBL}, macrophysics and shallow convection scheme. The Zhang-McFarlane deep convection scheme \citep{zhang:95} is still used. \ac{CAM}5-\ac{SE}/\ac{CLUBB} is shown with two different settings of the fourth-order horizontal diffusion coefficient. Figures~\ref{fig:aqua-CLUBB}(a,d) depict the default diffusion coefficient $1\times10^{15}$~m$^4$s$^{-1}$ for the $111$~km grid spacing (labeled $ne30np4$). The Figures~\ref{fig:aqua-CLUBB}(b,e) show the \ac{SE} results with an increased diffusion coefficient of $5\times10^{15}$~m$^4$s$^{-1}$. The SLD T85 dynamical core (Figures~\ref{fig:aqua-CLUBB}(c,f)) does not apply any explicitly-added diffusion since its numerical scheme already provides sufficient implicit numerical diffusion. The simulations shown in Figure~\ref{fig:aqua-CLUBB} are 1.5 years long and the first six months are discarded (model spin-up period). The physics and dynamics time steps for SE and SLD are the same as quoted before.

The top row of Figure~\ref{fig:aqua-CLUBB} shows the latitude-pressure cross section of a one-year-mean zonal-mean vertical pressure velocity in the tropics for SE (a,b) and SLD (c). The bottom row presents the one-year-mean zonal-mean precipitation rates of the three runs, split into total, large-scale and convective precipitation. The total precipitation rate can be directly compared to Figure~\ref{fig:aqua}.
Two observations are striking. First, the switch to the \ac{CLUBB} scheme causes the \ac{SE} and SLD dynamical cores with default diffusion settings (Figures~\ref{fig:aqua-CLUBB}(a,c,d,f)) to switch from the single \ac{ITCZ} shown in Figure~\ref{fig:aqua} to a double \ac{ITCZ} structure. Second, the appearance of a weak double \ac{ITCZ} structure in \ac{SE} (Figures~\ref{fig:aqua-CLUBB}(a,d)) is highly dependent on the choice of the horizontal diffusion coefficient. The increased diffusion coefficient in Figures~\ref{fig:aqua-CLUBB}(b,e) impacts the moisture processes in a way that convert the weak double \ac{ITCZ} in the default \ac{SE} run to a single \ac{ITCZ} peak. This brief assessment highlights the strength of an idealized testing framework in order to shed light on the physics-dynamics interactions. It is suggested that this approach can also be used to analyze the effects of different physics-dynamics coupling strategies.

\subsection{Summary}
The section demonstrated that simplified modeling frameworks and model hierarchies are useful tools to better understand
the differences between formulations on a lower level of complexity. This will simplify experimentation and interpretation. It was shown how the tests are sensitive to the different physics-dynamics coupling strategies, physical parameterizations and diffusion settings and their interaction with the dynamical core.
Here, simple prescribed SSTs (in aqua-planet mode) were used since the aim was to construct models that are as constrained as possible. Alternatively,
slab ocean models with predicted SSTs and a closed surface energy budget represent the logical next step in the model hierarchy.
In practice, models with significantly more complexity are utilized, with associated
physical parameterizations, data assimilation, and other infrastructure. Coupling these components together
(the same holds true for the land, ocean and ice models, chemistry, etc.) is non trivial, as the following section will describe.

\section{Intra model coupling}
\label{sec:intra}
In this section, the focus is on intra-models coupling problems within the climate 
modeling system, where the coupling occurs via an exchange of  boundary conditions that 
transmit fluxes through a physical interface (e.g. the ocean/atmosphere, land/atmosphere, ice/atmosphere or ocean/sea-ice interface). 
A difficulty inherent to this type of application is that many distinct 
physical processes at different temporal and spatial scales, governed by different 
physical/conservation laws, must be simultaneously considered as a whole. This difficulty 
leads to intertwined physical, mathematical and computational delicacies. 
%
Algorithms to solve such coupled problems can be classified into two general categories
\begin{enumerate}
\item[(i)] {\it Monolithic method}: a single model representing all components to 
be coupled is defined. It requires each component to share the same space-time 
computational grid and computational framework. The advantage is that a tightly (strongly) coupled
solution can be easily obtained. However, this approach is generally not tractable when trying to 
couple two individual models developed independently from each other with distinct numerical techniques, 
except for toy models \citep[e.g.][]{ganis11}. The monolithic approach has been used previously for land-atmosphere 
coupling when land surface processes were implemented as subroutines of \ac{GCMs} but is currently abandoned to provide 
more modularity because of the increasing complexity of land surface models which are now treated as external modules 
\citep[e.g.][]{polcher98, ryder16}. \par
\item[(ii)] {\it Partitioned/split method }: analogous to operator splitting, the full problem 
is split into smaller problems solved independently with boundary exchanges through their
common interfaces \citep[e.g.][]{schulz01, schmidt04, large06}. This is the most frequently adopted and most 
natural option in coupled problems arising in Earth system modeling. However, the difficulty is that this type of approach 
can give rise to various splitting errors and, thus, makes it difficult to recover a tightly coupled solution \citep{keyes13}. 
Analysis and attribution of these errors is not straightforward, as elaborated below. A comprehensive review about interface-coupled multiphysics 
systems in a broad sense can be found in \cite{keyes13}.
\end{enumerate}
Coupled problems arising in Earth system modeling cover a large range of aspects: 
parameterizations of turbulent boundary layers near interfaces, estimation of interfacial fluxes \citep{schmidt04, large06}, 
space-time numerical schemes, matching of different grids at the interface \citep[e.g.][]{best04, balaji07}, 
coupling algorithms \citep[e.g.][]{lemarie15, ryder16, beljaars16}, software implementation \citep[e.g.][]{valcke12}, etc, 
adding to the overall complexity of numerical models which are usually only considered on their own, neglecting connectivity. \par
In the present section the partitioned approach is considered and the example 
of the \ac{OA} coupling is used to illustrate the delicacies in terms of 
physics/dynamics inconsistency inherent to intra-model coupling. Most of the 
issues presented here are not only relevant to \ac{OA} coupling. Readers 
interested in more specific details on sea-ice/ocean or land surface/atmosphere 
coupling, for example,  could refer to \cite{schmidt04} or \cite{ryder16} and the references therein.\par

\subsection{Theoretical limitations of some of the current \ac{OA} coupling methods}
\label{sec:OA}
\begin{figure}
\noindent\includegraphics[]{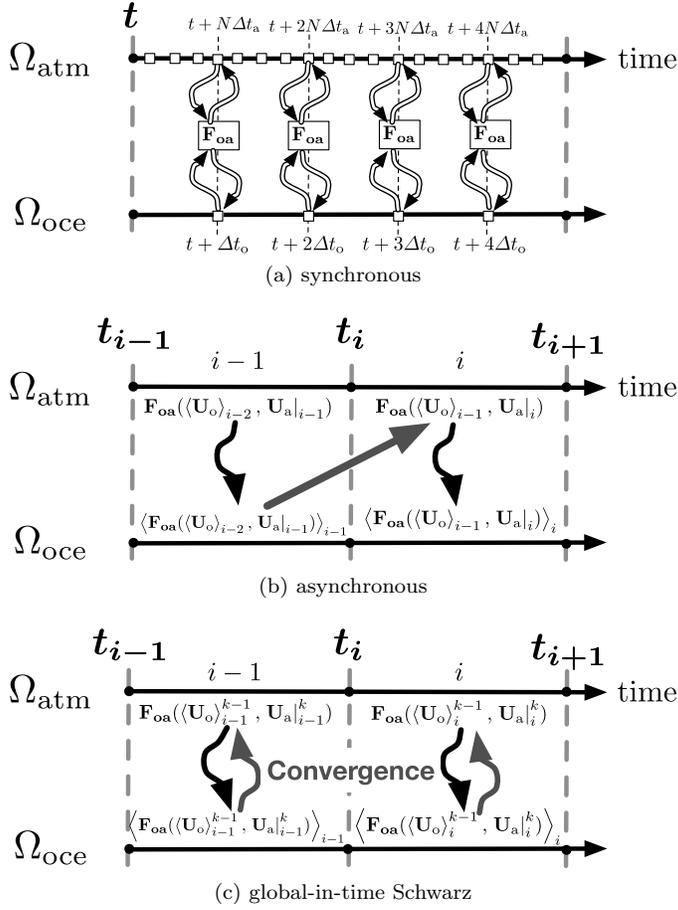}
\caption{Schematic view of the coupling between the computational
domains of the atmosphere model, $\Omega_{atm}$, and ocean model, $\Omega_{oce}$, with time advancing to the right. The function 
$\mathbf{F}_{\rm oa}(\mathbf{U}_{\rm o},\mathbf{U}_{\rm a})$ represents the parameterization of 
air-sea fluxes with $\mathbf{U}_{\rm o}$ (resp. $\mathbf{U}_{\rm a}$) the oceanic (resp. atmospheric) state vector. 
$\left< \cdot \right>$ is a given time averaging operator, and $\Delta t_{\rm o}$, $\Delta t_{\rm a}$ the dynamical 
time step of the models such that $N = \Delta t_{\rm o}/\Delta t_{\rm a}$.}
\label{fig:algos}
\end{figure}

Most multiphysics coupling problems assume that all scales are resolved by the numerical 
models and that the boundary conditions at the interface are of Dirichlet or Neumann type (or a linear 
combination of both). In the case of the ocean-atmosphere problem the dynamical coupling is strongly 
influenced by physical parameterizations which makes the rigorous mathematical analysis not tractable.
The numerical resolution of the \ac{OA} coupling problem for practical applications is generally tackled in two different ways. Either by an exchange of instantaneous 
boundary data at the largest time step of the two models, 
this method is referred to as synchronous coupling, or by an exchange of averaged-in-time boundary 
data over a time interval $[t_i,t_{i+1}]$ (which is much larger than the largest time step). The latter method is referred 
to as asynchronous coupling. Those methods, described in Figure~\ref{fig:algos}, are loose coupling schemes (in contrast to tight coupling schemes)
in the sense that they correspond to only one iteration of an iterative process without 
reaching convergence \citep[see][]{lemarie14, lemarie15}. Hence, they do not strictly provide the solution to the \ac{OA} coupling 
problem, but an approximation of one since state variables of the two models are shifted by one time-step or a sequence of time-steps. 
The theoretical limitations of the synchronous and asynchronous methods are now explained further. 
In the synchronous coupling algorithm, the following errors are observed:
\begin{itemize}
\item {\it Aliasing errors}: significantly different time steps are used in each model (for 
the same horizontal resolution the oceanic model is integrated with a time step 
approximately ten times larger than the atmospheric model), as a consequence aliasing problems may arise 
and compromise stability \citep[e.g.][]{schluter05}
\item {\it Synchronicity error} : air-sea fluxes are used as boundary conditions for the vertical turbulent diffusion 
terms which are treated implicitly in time, meaning that the fluxes at the interface are formally needed at time 
$t +\Delta t$ and not $t$ (in Figure~\ref{fig:algos}a). The explicit exchange of data in the 
synchronous coupling leads to an additional condition for the coupling to be stable even if unconditionally 
stable time stepping algorithms are used for vertical diffusion \citep{lemarie15, beljaars16}.
A way to circumvent those stability issues is to consider a synchronous coupling with an implicit exchange of date. 
In practice, this approach amounts to solve the local implicit problems in the ocean and the atmosphere in a monolithic way 
as one single implicit solver as often done for land-surface/atmosphere coupling \citep{polcher98, ryder16}. Implicit flux 
coupling is so far seldom used in the context of \ac{OA} coupled models.
\item {\it Physics-dynamics inconsistency error}: the uncertainties in the computation of air-sea fluxes at high-frequency 
through bulk formulations is extremely large (see discussion in section~2 in \cite{large06}, or \cite{foken06}). The sources 
of those uncertainties are numerous, among them are the assumptions used to derive the continuous formulation of bulk 
formulae (constant-flux layer assumption, horizontal homogeneity, quasi stationarity, etc) and the fact that very few direct 
measurements exist to calibrate those semi-empirical formulations over the ocean. Moreover, the nonlinear problem associated 
to the estimation of bulk fluxes is often solved in an approximate way. In practice an averaging of the oceanic and atmospheric 
inputs to the bulk formulae should be required to minimize the uncertainty in air-sea fluxes \citep{large06} meaning that an internally 
required time-scale $\Delta t_{\rm phys,req}$  needs to be assumed for the parameterization scheme (a.k.a bulk formulation) to be 
valid, and $\Delta t_{\rm phys,req}$ is generally larger than the model dynamical time step $\Delta t_{\rm dyn}$. 
As a result, using a synchronous method can render the model solution very sensitive to the choice for the time step 
$\Delta t_{\rm dyn}$ since it is implicitly assumed that $\Delta t_{\rm phys} = \min(\Delta t_{\rm dyn,req},\Delta t_{\rm dyn})$, 
which can lead to large errors in the estimation of air-sea fluxes. 
\end{itemize}
By construction the asynchronous coupling is expected to mitigate this latter issue since boundary data 
averaged-in-time are exchanged over a time interval $[t_i,t_{i+1}]$ generally much larger than the dynamical 
time-step. However, the asynchronous coupling algorithm also suffers from a synchronicity issue.
Indeed, the oceanic state used on $[t_i,t_{i+1}]$ comes from the previous time window $[t_{i-1},t_{i}]$ 
and not the current time window. Note that the lack of synchronicity is clearly visible in Figure~\ref{fig:algos}b 
(oblique arrow). This error arises from the use of a non-iterative partitioned coupling approach. The asynchronous coupling does 
not permit an accurate representation of transient processes on short time scales, e.g., the diurnal SST cycle, 
which is undesirable especially when the space-time resolution is increased.
This approach is however still used in numerous climate coupled models but an active research is currently 
in progress to minimize those synchronicity issues and allow correct phasing between the ocean and the atmosphere 
at a reasonable computational cost. 
\subsection{Reducing physics-dynamics inconsistency and splitting errors}
In order to explore possible ways to reduce the above mentioned errors 
the theoretical framework of the Schwarz-like domain decomposition 
methods \cite[e.g.][]{lemarie13} can be adopted. The idea behind those methods is to separate 
the original problem on the computational
domain $\Omega$, with $\Omega = \Omega_{\rm oce} \cup \Omega_{\rm atm}$
into subproblems on 
$ \Omega_{\rm oce}$ and $\Omega_{\rm atm}$, which can be solved separately. An iterative process 
is then applied to achieve convergence to the solution of the original problem. 
For the coupling of systems of partial differential equations (ignoring physical parameterizations) the 
converged solution obtained using the Schwarz algorithm is the same as the one 
obtained using a monolithic approach, within a given tolerance. It can be shown 
that the asynchronous coupling method  corresponds to a 
single iteration of a global-in-time Schwarz Method, see Figure~\ref{fig:algos}c. 
Respectively the synchronous coupling method  corresponds to one 
single iteration of a local-in-time Schwarz method.

The usual methods (e.g. synchronous and asynchronous coupling) used in the context of ocean-atmosphere coupling are thus prone 
to splitting errors because they correspond to only one iteration of an iterative 
process without reaching convergence. There are so far very few studies aiming at 
quantifying the impact of these coupling errors on the coupled solutions. In \cite{ganis11}, 
using highly simplified models, it is shown that the use of a synchronous algorithm 
compared to a monolithic approach (or equivalently a Schwarz algorithm) leads to 
a larger model uncertainty in the sense that the resulting variance in model variables 
is larger. This result is based on an uncertainty quantification method using stochastic 
input parameters for the exchange coefficients involved in the air-sea flux computation. 
In \cite{lemarie14}, numerical experiments using a mesoscale atmospheric model 
\ac{WRF} coupled with a regional oceanic model \ac{ROMS} for a realistic simulation of 
a tropical cyclone have been carried out. Ensemble simulations have been designed 
by perturbations of the coupling frequency and the initial conditions. One ensemble 
has been integrated using the global-in-time Schwarz Method and an other using the 
asynchronous method. The Schwarz iterative coupling methods leads to a significantly 
reduced spread in the ensemble results (in terms of cyclone trajectory and intensity),
thus suggesting that a source of error is removed with respect to the asynchronous coupling case.
The results of \cite{ganis11} and \cite{lemarie14} emphasize empirically a correlation 
between the existence of splitting errors and model uncertainties.\par
Physics/dynamics inconsistencies in the context of coupled problems are hard to estimate 
since there is a lack of academic test-cases with reference solutions including 
physical parameterizations. It has been shown earlier that such inconsistencies can 
arise from coupling algorithms or nonconformities in the space-time computational grids 
but also from parameterization schemes for air-sea fluxes and turbulent boundary layers.
However, the mathematical formulation of those schemes is often devised semi-empirically (e.g. by fitting 
independent measurements) and this can impair the regularity of the associated solutions 
\citep[e.g.][]{burchard05,deleersnijder08}, giving rise to the development and persistence 
of "fibrillations" in model solutions. This complexity has to be taken into account when  
designing mathematically consistent and efficient intra-model coupling algorithms. Indeed the use of an 
iterative coupling method requires a certain degree of regularity (well-posedness) of 
the system of equations to be coupled, otherwise convergence is not guaranteed and/or 
the relevance of the converged solution could be questionable. For instance, the theoretical framework 
of the Schwarz methods could be used to derive intra-model compatibility/consistency constraints on the 
turbulent boundary layers parameterizations: a pair of parameterizations will be declared 
compatible if the associated iterative Schwarz algorithm converges. This task is challenging 
because parameterization schemes are often very complicated.
To investigate those issues, working on simplified equation sets to focus on specific 
problems should be encouraged. For example, coupling in a \acp{SCM} is 
representative of the functioning of the three-dimensional coupled models 
(see \cite{schulz01} for an example in land surface/atmosphere coupling). Because 
one-dimensional coupled \ac{LES} are increasingly used, they should provide 
reference solutions to coupled \acp{SCM} for various physical situations. \par

As emphasized above, the coupling of models can only succeed if the problems themselves are well posed, the coupling schemes sufficiently advanced to allow for convergence, and, crucially, the processes in different models are compatible. 
This can only be the case if their thermodynamic basis is compatible, otherwise no common state can exist (unless be coincidence). 
The aspect of the thermodynamic formulation will be explored in the following section on a example in the atmosphere.

\section{Thermodynamics / inter process coupling\label{sec:thermod}}

\subsection{Inclusion of moisture related quantities}
With increasing mesh resolution, the impacts of cloud physics processes become more and more variable. More extreme precipitation events become explicitly  represented. Associated with such events there can be large, but transient, amounts of liquid water in a single model column (\cite{GeleynMarquet11,Bacmeisteretal12}).
From a physical viewpoint thermodynamic properties such as density, the gas constant or the specific heat of the air-cloud-precipitation mixture may no longer be approximated by the properties for dry air. Also the barycentric velocity $\mathbf{v}=\Sigma_i\mathbf{v_i}\varrho_i/\varrho$ (with $i=$ dry air, water vapor, rain, snow, etc) may differ from the velocity of dry air  when a significant amount of condensate  is present.  Those issues are especially important as the largest amount of available energy in the atmosphere is processed by condensation/evaporation processes and precipitation. It has also been found that cloud physics is responsible for most of the variability of regional extreme event forcasts~\citep{GeleynMarquet11,Bacmeisteretal12}, e.g. flash floods. Non-convergence in any model is largely driven by the different behavior of cloud physics  as the time step is changed (see section~\ref{sec:arti}). The single microphysical conversion rates, sedimentation rates, and turbulent or convective flux amounts are not exactly known. The only physical constraint available is correct energy transformation of all moisture related processes. The sole constraints of modeling are total mass conservation and total energy conservation. Those conservation properties on the global scale are generated by local phase changes and local latent heating rates. Therefore the expected changes from the current and often quite loose approximations to more physically consistent modeling will first appear on local scales and will have impact on the simulation of localized events such as flash floods due to extreme precipitation. 

Another modeling constraint is interrelated tracer consistency (e.g. \citep{QJ:QJ986}. If reactive gases are not transported coherently, spurious micro-physical or chemical reactions are the consequence. Such kind of inconsistencies will in turn also affect the radiation scheme if for instance ozone chemistry is explicitly modeled.

To some extent, the mass conservation constraint and the interrelated tracer consistency constraint do not go hand in hand. The modeling groups  have two choices: either using specific quantities $q_i=\varrho_i/\sum_i\varrho_i$ or mixing ratios $r_i=\varrho_i/\varrho_d$, where $\varrho_d$ is the density of dry air, as prognostic variables. The choice of specific quantities goes hand in hand with classical thermodynamic theory known from textbooks (c.f. \citep{deGrootMazur84}). This choice supports energy and mass conservation and leads easily to the derivation of a meaningful entropy budget equation. It is also consistent with the mass-weighted averages that are assumed in turbulence modeling. A drawback of specific quantities is that the dry air compartment is not a prognostic variable, but its evolution is determined by the continuity equation for total mass. The specific dry air content is then derived from $q_d=1-\sum_i q_i$, where $i$ denotes all other constituents but dry air. Therefore, an interrelated tracer consistency between dry air and other constituents is numerically no longer possible. For interrelated tracer consistency, the choice of mixing ratios is more adequate. Then the first and second laws of thermodynamics have to be re-derived adequately, which may lead to other problems.

One of the fundamental assumptions for modeling multi-phase flows in the atmosphere is a constant temperature throughout the air parcel. Hence, falling precipitation is always assumed as having the same temperature as its environment. This assumption is very crude, but relaxing it would multiply the number of prognostic variables.

Only recently have the influences of correct thermodynamics been investigated. Although correct governing  equation sets including all moisture effects have been known for some time
(\cite{Bannon02,ZdunkowskiBott03,WackerHerbert03,Wackeretal06,Catryetal07,GassmannHerzog15}),
a lot of models still approximate thermodynamics severely without the consequences being well understood. The PDC workshop  series, \cite{Gross2015}, intends to provide a platform for bringing attention to this issue.

Consistent thermodynamics are also needed to improve the understanding of the static stability of the atmosphere, which is important when diagnosing the intensity of turbulence. For this purpose, moist entropy, moist potential temperature and moist Brunt-V\"{a}is\"{a}l\"{a} frequency have been derived by Marquet and Geleyn in a sequence of publications
(e.g. \cite{MarquetGeleyn13}).

\subsection{Compatibility with the second law of thermodynamics}
\begin{figure}
\begin{center}
\includegraphics[width=15cm]{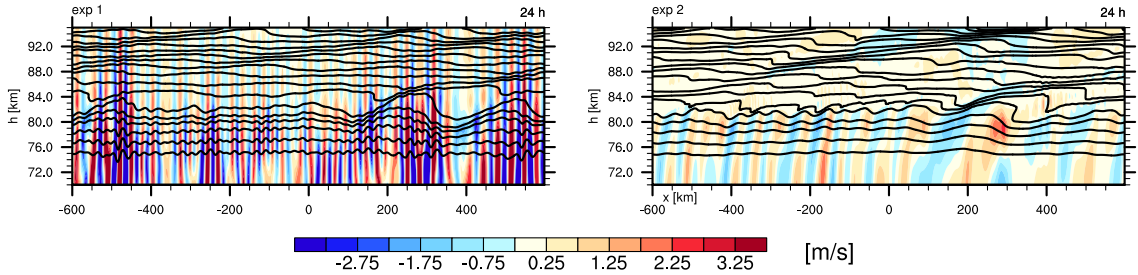}
\caption{\label{fig:gassmann_w24std} Vertical velocity $w$ (colors) and selected isentropes (contours) for  the late phase of gravity waves breaking in the mesosphere for a  classical turbulent heat flux parameterization (left) and a 2nd-law  compliant variant (right). For this idealized 2-dimentional slice  experiment, gravity waves are excited in the troposphere during the  first 16 hours. They travel through an idealized summer stratosphere.  The flow field after 24 hours is shown.}
\end{center}
\end{figure}

Surprisingly, modeling communities have not yet paid much attention to the compatibility of the model formulation with the second law of thermodynamics \citep{GassmannHerzog15}.
The currently widely accepted argument supporting this disregards that turbulent fluxes are not considered structurally similar to molecular or viscous fluxes, which are indeed responsible for irreversible processes in nature \citep{deGrootMazur84}. On the other hand, as already mentioned, the r\^{o}le of the physics parameterization is the irreversible conversion of some kind of available (kinetic or internal) energy into unavailable internal energy, whereas the dynamics is responsible for the two-way energy transfer. Since there is no other means than the dynamics for reversible energy conversions, the physics has to follow structurally the same rules as known from viscous or molecular processes. Therefore it is straightforward to raise the question of entropy production in the classical sense also for a numerically modeled atmosphere, as done for instance by  \cite{Goody00,doi:10.1175/JAS-D-14-0361.1,doi:10.1175/JAS-D-16-0240.1,PauluisHeld02} and \cite{Romps08}. The last three authors recognize a serious problem: Turbulent heat fluxes proportional to the negative gradient of potential temperature give rise to negative entropy production and offend the second law if the atmosphere is stably stratified and turbulence is mechanically driven, as it is the case for gravity wave breaking. In this case, work has to be performed to push isentropes down in subgrid turbulent parameterizations. However, this work is not visible in the kinetic energy budget of the whole model. In order to solve the problem, \cite{Akmaev08} claims that the frictional heating -- or shear production of turbulent kinetic energy -- provides the necessary energy. However, from a process-oriented thermodynamic viewpoint, turbulent momentum fluxes and turbulent heat fluxes are unrelated processes according to Curie's principle, because the tensorial character of both fluxes is different \citep{deGrootMazur84}. Current and yet unpublished research (Gassmann, in preparation) aims at the resolution of the problem by introducing a turbulence-related vertical pressure gradient term in case of stable stratification. Then, the work done by buoyancy forces to push isentropes down becomes obvious as a loss of the vertical part of the resolved specific kinetic energy of vertical motion ($w^2/2$). The newly introduced turbulence-related vertical pressure gradient term acts like a Rayleigh damping term in the vertical velocity equation. Such kind of Rayleigh damping has already been used as a numerical remedy filtering gravity waves in a sponge layer of an atmospheric model by \cite{Klempetal08}. The vertical diffusion coefficient for potential temperature which is assigned to this apparent Rayleigh damping coefficient will have other properties than current diffusion coefficients. The diffusion coefficient depends on the kinetic energy of vertical motion. Therefore, isentropes are not pushed further down at their trough position and breaking waves can not amplify as would be the case in a conventional parameterization. Figure~\ref{fig:gassmann_w24std} demonstrates that a classical 2nd-law offending parameterization can lead to wave manifestation and amplification, but the 2nd-law compliant parameterization leads to wave attenuation. 

If the turbulence is convectively driven, the vertical heat flux might be counter the local gradient of potential temperature in the upper part of the convective layer. The energy supply for upward mixing originates from the \ac{TKE}. From the perspective of the resolved model variables, the \ac{TKE} and the internal energy are indistinguishable and are only representable together as internal energy. Unlike in the mechanically driven turbulence described above, an energy exchange between resolved kinetic energy and internal energy does not take place. Therefore the model equations must pretend a downgradient temperature (heat) flux $c_p\bar\varrho\overline{w''T''}=-c_p\bar\varrho K^T\partial_z\bar{T}$. Such a flux is always compliant with the 2nd law of thermodynamics for positive $K^T$. From the phenomenological side which knows about the distinction between \ac{TKE} and internal energy, a flux formulation according to $c_p\bar\varrho\overline{w''T''}=-c_p\bar\varrho(\bar{T}/{\bar\theta}) K^{\theta}(\partial_z\bar\theta-\gamma)$ is adequate. The countergradient term $\gamma$ does not lead to a violation of the second law as long as the total heat flux is down the true temperature gradient. In other words: for given $K^{\theta}$ and $\gamma$ from an arbitrary parameterization, the 2nd law of thermodynamics requires from the relation $-c_p\bar\varrho K^T\partial_z\bar{T}=-c_p\bar\varrho(\bar{T}/{\bar\theta}) K^{\theta}(\partial_z\theta-\gamma)$ that $K^T$ is a positive number. What is usually referred to as a counter-gradient flux is therefore not always violating the second law.

The discussed heat fluxes for stable and unstable stratification are the most intriguing examples where current understanding and practice has not yet matured. 
\cite{GassmannHerzog15} derive a slightly different approach for turbulent fluxes of water vapor and include a symmetric and trace-free tensor depending on shear and strain deformations in the turbulent momentum diffusion .

An issue which also becomes important with regard to the second law is the not well understood role of numerical diffusion. A lot of
recent dynamical core developments have focused attention on geometrically compatible formulations, i.e. they require that some underlying geometrical law, such as the duality of divergence and gradient formulations or the curl-free gradient operator, hold in a discrete sense. Formulations that strictly respect this form of geometric compatibility lead to dynamical cores with truly reversible energy transformations, so that transformations between different kinds of energy are equally possible in both directions. Some diffusion in dynamical cores is still motivated numerically, even though this interferes with physically motivated turbulence parameterization schemes. Future work should consider the relation of these two effects (numerical diffusion / turbulence parameterization) and try to interpret numerical diffusion in a physical sense.

\subsection{Future research}

A question posed to the modeling communities is: Should all the parameterizations obey the second law? If the answer is  yes then the consequence is that convection and gravity wave drag parameterization schemes have to be interpreted  as special cases of ordinary subgrid-scale turbulence parameterizations. If the answer is  no then are the efforts put into the developments of geometrically compatible dynamical cores worthwhile? Parameterization schemes of the \ac{EDMF} type  would then partly perform functions of the dynamical core, because they treat advection and diffusion at the same time.

\section{ The gray zone }
\label{sec:grey}
As the resolution of geophysical models increases, the scale
separation between the scales which are resolved by the model and the
scales of the subgrid processes which are parameterized is vanishing.
When a subgrid process, which is targeted by parameterizations at a
lower resolution, is becoming partially resolved at a higher resolution,
the model runs at a resolution in what is called the {\em gray zone} of
this process.

\subsection{Examples in current model configurations}
The continuous resolution increase over the last 50 years has brought
global models very close to the gray zone of convection. Limited area
models for \ac{NWP} have already jumped across the
gray zone of convection, but they are now at the verge of the gray
zone of turbulence \citep{Wyngaard2004,Boutle2014,Honnert2014}.

In the gray zone of a process, neither the subgrid representation of
this process in a parameterization, nor the explicit representation by
the prognostic parameters of the model are designed to accurately
represent this process.  Often, in practice, modelers either tune the
existing parameterizations to extend their usage out of the range of
validity of their fundamental hypotheses, or they switch off the
parameterizations, even if the process is not yet well resolved but
only ``permitted'' by the resolution, applying ad-hoc numerical
filters, if necessary, to control the intensity of the process.  

\begin{figure}
\begin{center}
\includegraphics[]{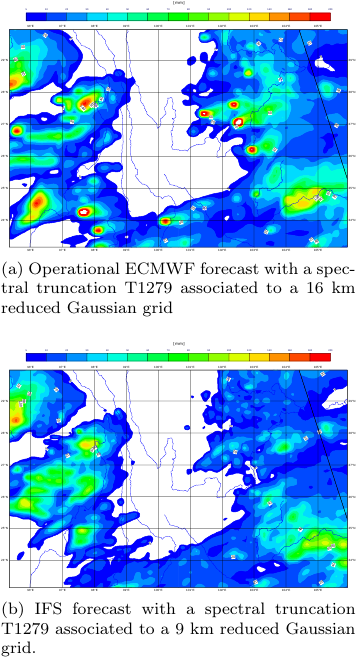}
\caption{\label{rainbomb} 3 day accumulated surface large scale precipitation for forecasts starting on 2015-05-20 at 00~UTC VT:23 May 2015 00 UTC.}
\end{center}
\end{figure}

A recent example in the operational \ac{ECMWF} model showed the difficulty of balancing the
explicit and the parameterized representations of deep convection, even
at a resolution of about 16~km which is not considered to be in a gray
zone yet. When the convection scheme of the \ac{IFS} was modified to improve the daily cycle of
convection \citep{Bechtold2014}, explicit convective clouds at
isolated single grid points started to be diagnosed in calm conditions
near mountainous and moist areas, leading locally to very unrealistic
precipitation (Figure~\ref{rainbomb}(a)). This new version of the
convection scheme then delayed the onset of the parameterized convection
toward the evening. The \ac{CAPE}
accumulates such that, in a region with weak orographic forcing of
moist air in a no shear environment, an explicit convective cloud is
permitted at a single grid point before the convection scheme is
triggered. In the \ac{IFS}, such single grid point structures are then
pathologically amplified by the semi-Lagrangian advection scheme of
the \ac{IFS} dynamics \citep{QJ:QJ2509}. The resulting unrealistic explicit
deep convective clouds may last for several hours.  This leads to
spurious and very high precipitation rates at some grid points.
Similar grid point storms have been reported in the literature for
other global models, for example \cite{QJ:QJ1992} in \ac{CAM}4, with T340 spectral truncation and a 5 min
time step, (c.f. section~\ref{sec:arti}) or in mesoscale limited area models at
resolution in the 5-3~km range \citep{QJ:QJ2509}.
This example shows that, even with a hydrostatic model such as the \ac{IFS}, explicit deep convective circulations are permitted at resolutions which are far too coarse to sample individual convective ascents in circumstances where the convection scheme is not triggered soon enough to release the \ac{CAPE}. On the other hand, with the old version of the convection scheme, the parameterization was triggered earlier, and then the onset of the convection in the tropics was systematically too early. Finding the right balance between the explicit and the parameterized representation of convection everywhere around the globe becomes even more difficult in the gray zone of convection, at resolution around 5~km. 

In the next sections, the main limitations of atmospheric models in the gray zone of convection and turbulence are listed. Recent attempts to design algorithms improving the seamless transition between resolved and parameterized treatments in these gray zones are presented.
\subsection{Model limitations in the gray zones}
\label{sec:modlelimgrey}
In a gray zone, the explicit and the parameterized representations of a process are in ``competition'' in the numerical model. The result of this
competition may be double counting or no counting at all, or, more
often, the process is taken into account by both options but
in a sub-optimal manner.

The prognostic equations which are solved by a numerical discrete model are the result of time and space filtering. This filtering creates an artificial cut-off in the
continuous atmospheric spectrum between the processes which are represented by the mean prognostic variables of the models and the processes which are supposed to be ``subgrid'' and whose effect on the larger scales is parameterized.
This cut-off scale is partly defined by the time/space
resolution of the model and partly by the characteristics
of the numerical schemes and physical parameterizations.
In most numerical schemes, the
largest errors are expected to happen at the cut-off scale, especially in
regions of large gradients. Weaknesses of the numerical schemes such as large diffusion, large
phase shift or non-conservation  then directly affect the energy-containing
circulation if they are permitted at this cut-off scale.
Thus, as discussed by \cite{Lander1997}, physical parameterization should not force and should not be forced by the prognostic model variables containing variance at the grid scale. This statement is actually well illustrated with the case shown in Figure~\ref{rainbomb}. With the horizontal resolution upgrade at the beginning of 2016, the \ac{IFS} moved from a "linear" grid to a "cubic" grid, keeping the same spectral truncation T1279. With the cubic grid, the smallest wavelength of the truncation is now represented by 4 points instead of 2 with the linear grid. Such a pairing between the spectral representation of the \ac{IFS} and the model grid insure a better scale separation between the prognostic model variables and the subgrid effect computed in the physics package. In particular, single grid column resolved ascents are not allowed any more with a cubic grid. The development of grid point storms is then completely eliminated from the \ac{IFS} forecast (Figure~\ref{rainbomb}(b)). 

Indeed, the formulation of most physical parameterizations is based on the clear
scale separation, both in time and space, between a resolved environment and the
parameterized processes which are treated as ``perturbations'' of the environment.
The formulations are derived from a statistical evaluation of the
impact of a large population of ``perturbations'' on the resolved
flow, sometimes simplified by a ``bulk'' representation of the
process, as, for example, in many convection schemes, a single
convective cloud replaces a population of smaller cloud ascents.  But,
when the resolution increases, the grid becomes too small for a
population of deep convective circulations to develop inside a grid box. 
In the gray zone of convection, the updrafts could
cover a large fraction of the grid box area, and then the mean grid-box
properties which are carried by the prognostic variables of the model should depart substantially from the updraft environment.
The detrained material from the updrafts should also not be confined
in the same grid column, but should be distributed over several grid
columns  (cf. section~\ref{sec:simplified_eq_1} and its Figure~\ref{fig:seq1}).

However, the Reynolds decomposition which is used
to derive the eddy diffusivity formulation for non-local 1D turbulence
or the mass flux formulation of most convection schemes does not allow
any net mass transport by the ``perturbations'' in a grid box: updraft
and downdraft have to cancel in the same grid box, 
rendering the problem of extending the detrainment to neighboring cells difficult to generalize (c.f. section~\ref{sec:simplified_eq_1}).

The equilibrium hypothesis \citep{Arakawa1974} resulting from
the hypothesis of the scale separation in time also starts to collapse
when the time step of the model decreases. With an increase of
resolution in time, the variability of the model variables becomes
faster than the characteristic convective time scales, \cite{Gerard05}. If the closure of the convection scheme is too simple,
spurious explicit convective storms are more likely to develop at high
resolution \citep{QJ:QJ1992,Gerard2015} (cf. section~\ref{sec:splittingissues}).

The generalization from a 1D to a 3D treatment of a subgrid process often does not only involve the physical concepts behind the parameterization but also the code architecture of the model.
Indeed, the scalability of the model benefits from the independence between the columns in the physics. Thus, the transition from 1D algorithms to 3D parameterizations involving horizontal exchanges with neighboring columns is also a challenge in term of code development and efficiency of the algorithms, especially for \ac{NWP} and climate models.

\subsection{Towards scale aware parameterizations}
Efforts to develop scale-aware convective parameterization started in
the \ac{LAM} community more than ten years ago \citep{Gerard05} and are
now shared by a much larger community \citep{Arak13,Gustafson2013,Grell2014,Sieb15}.

As already pointed out in the previous section, one of the main issues for the parameterization of deep convection in the gray zone is that, when the resolution of the model increases, some of the condensates can be detrained across the gridbox and the ensuing compensating subsidence should take place within another grid box than the originating one. With the time step organization of \ac{NWP} codes, this transport can only be handled by the advection of the dynamical core.

\cite{Piri07} observed that the advantage of  \acp{CSRM} with respect to parameterized budget equations is that the source terms for the convection can be separated into transport terms and microphysic terms and they argued that the two types can be treated independently. Moreover, if the condensation (and the cloudy evaporation) terms in cloud budgets models are computed by a microphysics scheme and provided as source terms to the environment, then the system can be closed, leading to \ac{CSRM}-type equations that still do not contain detrainment terms. In that case, there is no need to directly rely on the budget equations to close the system.

However, to go from \acp{CSRM} to gridbox parameterizations, 
it is necessary to partition the grid box into a convective and a non convective part.  \cite{Gerard09}  used the cloud scheme of  \cite{Xu96} and \cite{Smith90} to introduce a protection of the cloud condensates in the convective part, to prevent their evaporation by the cloud microphysics scheme.
Additionally they used a prognostic formulation of the convective mesh fraction of the updraft and a prognostic equation for the updraft-vertical velocity proposed in \cite{Gerard05}. The result is a \ac{CSRM}-type set of equations without any explicit presence of detrainment terms. In other words, it interacts with the dynamics in the same manner as a \ac{CSRM}-type of model does. 

One can argue that bulk parameterizations should converge in their behavior to the behavior of \acp{CSRM} in the cloud-resolving limiting resolutions.  If the prognostic equations of the mesh fraction and the updraft-vertical velocity scale properly, then the equations should converge to the equations of a \ac{CSRM}. This yields a mechanism to control this convergence and to formulate a scale-aware parameterization of deep convection. 
	
This approach was implemented in a scheme called the \ac{3MT} and it formed the basis of the so-called ALARO-0 configuration of the ARPEGE-\ac{ALADIN} system. \cite{Gerard09} showed satisfactory results of this scheme with resolution ranging from the mesoscale up to 4~km, cf. Figure~11 of \cite{Gerard09}, where it can be seen that without \ac{3MT} the model did not resolve the organized convection satisfactorily, resolving only a small (in number and size) of intense gridpoint storms.

Recently, good results were found with an updated 
version of the scheme up to a resolution of about 1~km. \cite{DeMeu15} tested a 
version of the 3MT scheme which includes the parameterization of unsaturated downdrafts. They found downdraft mass fluxes that are sufficiently realistic so that they can be used operationally to forecast downbursts. \cite{DeTro13} demonstrated that the ALARO model has an improved multiscale character than the former \ac{ALADIN} configurations.


These efforts still need to be generalized for global NWP. There very different types of convective
circulations, in the tropics and at higher latitudes, have to be well
represented for medium range weather forecasting. It is also of importance for climate models in order to
maintain a correct large scale balance. Recent
results with variable resolution meshes \citep{Mul14} also show the
need for scale-aware physics across the difficult range between
$1-10$~km.

LAM models which are running at sub-kilometer resolutions do not use any parameterization of deep convection. But, the parameterization of eddies in the boundary layer is still needed. A blending between a 3D turbulence parameterization originally designed for \ac{LES} and a 1D boundary layer parameterization suitable for coarser grid resolution has been shown to be very beneficial to the representation of clear or stratocumulus-topped boundary layers by \cite{Boutle2014}. In this case, the transition laws brought out by \cite{Honnert2014} are used to seamlessly drive the transition between unresolved to resolved turbulence. But more efforts are needed to generalize this approach to any regime in the boundary layer.

\cite{Mal15} identify the influence of subgrid scale
parameterizations for the shape of the  \ac{KE} spectra as well as for the
non-linear spectral fluxes at all scales.
The artificial scale separation between resolved and subgrid processes
modifies the natural turbulent energy cascade driven by advection.
When the processes are parameterized, the circulation which is responsible for
the average effect of the subgrid mixing is neither part of the
``resolved'' \ac{KE} spectra, nor part of the non-linear spectral
transfer, thus effectively disabling any energy cascade. 

The temptation to enable the natural cascade by eliminating a
particular parameterization too early in a gray zone is however risky
as the model balances change at all scales as a result. Such practice
may also have implications on the forecast error growth, as the
predictability time of a $k^{-3}$ system can be much longer than that
of a $k^{-5/3}$ system \citep{Palm14}. However, it is unclear if
a growing error is merely replaced by a much larger error injected at
multiple scales when the process is parameterized.

The continuous increase of resolution is not the only change in the model environments that generate \ac{PDC} challenges. Variable resolution and high order methods for example are two further areas that emerge on the horizon and are discussed in the following section.

\section{ Emerging challenges }
\label{sec:emerging}
The ecosystem of models is constantly evolving and new methods become available and feasible, replacing older, often somewhat simpler technologies. Currently the advent of high order finite element methods offers many more choices to the coupling than a grid point model would. Likewise spatially varying resolution and or adaptive refinement is used more and more often, partly due to the availability of mimetic methods which support this sort of models. This however is not without challenges to the coupling of the multiple (truncation) scales now present in the model. 

\subsection{Spatial physics-dynamics coupling with element-based high-order Galerkin methods}
\label{sec:emerging_fe}
Numerical methods based on element-based high-order Galerkin discretization \citep[see, e.g., ][]{Durran} have reached a level of maturity in which they are being considered for next generation weather and/or climate models. For example, the spectral-element dynamical core in \ac{NCAR}'s  \citep[\ac{CAM}; ][]{CAM5}, referred to as \ac{CAM}-\ac{SE} \citep{TES2008JPCS,TF2010JCP,DetAl2012IJHPCA}, is currently being used for high resolution climate modeling \citep[e.g., ][]{JAME:JAME20125}. Other examples are \citet{Giraldo20083849}, \citet{Nair2009309} and \citet{BSBDK2013TCFD}. Below the focus is on \ac{CAM}-\ac{SE}, however, in principle the discussion applies to any element-based high-order Galerkin method.

To advance the solution to the equations of motion in time, element-based high-order Galerkin methods typically apply quadrature rules to numerically integrate basis functions over a reference element. The choice of quadrature rule is application dependent and can have consequences for the properties of the final algorithm; in particular algorithm efficiency. In \ac{CAM}-\ac{SE}  \ac{GLL} quadrature is used which exactly integrates Lagrange polynomials up to degree $2p-1$, where $p+1$ is the number of quadrature points. For an introductory discussion on emerging Galerkin methods in the context of atmosphere modeling see, e.g., \citet{NLL2011LNCSE}.

Irrespective of the choice of quadrature rule, the quadrature points for higher-order methods are not equally spaced over the sphere and reference element. The higher the order, the more the quadrature points tend to cluster near the sides and, in particular, corners of the elements. 
 As far as the authors are aware, current dynamical cores employing element-based high-order Galerkin methods use the quadrature point values for the state of the atmosphere passed to subgrid-scale parameterizations (physics). This approach follows the traditional model setup where physics and dynamics grids coincide. One may question if that is an appropriate choice for element-based high-order Galerkin methods. Physical parameterizations expect a state of the atmosphere representative of the area for which it should compute tendencies, a grid-cell averaged state of the atmosphere, for example. Obviously the quadrature point values are representative of the state of the atmosphere at the quadrature point and in the vicinity of the quadrature point but what area is associated with the quadrature point value? If one defines areas around the quadrature points so that the spherical area exactly matches the \ac{GLL} quadrature point weight times the metric factor, a very irregular grid results. 
Hence the state of the atmosphere passed to physics is sampled anisotropically in space. 

 \begin{figure}
 \noindent\includegraphics[width=16pc]{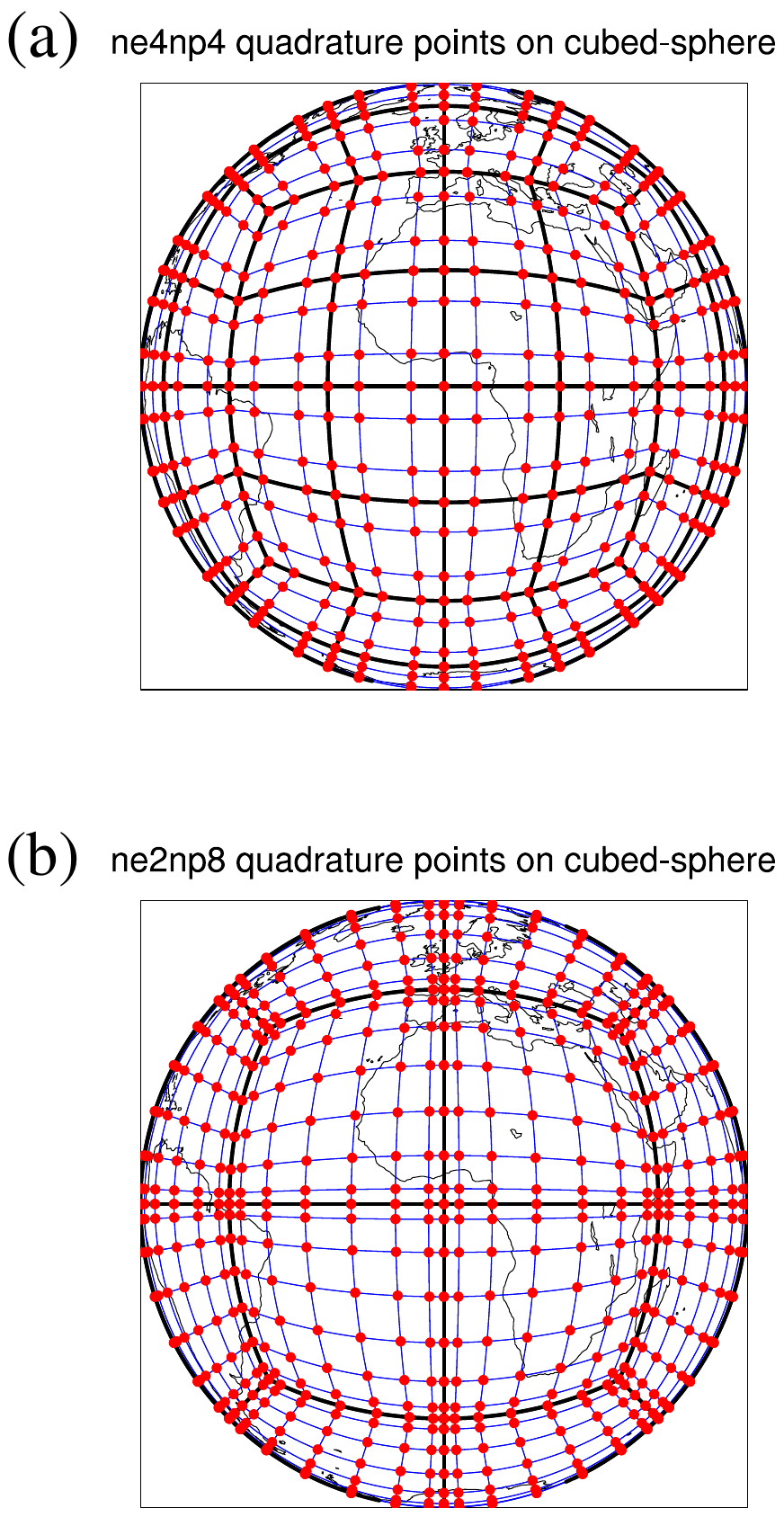}
 \caption{The \ac{CAM}-\ac{SE} \ac{GLL} quadrature grid (red filled circles) on the cubed-sphere using (a) $4\times 4$ elements per panel ($ne=4$) and $4\times 4$ quadrature points in each element ($np=4$), referred to as $ne4np4$ grid, and (b) $2\times 2$ elements on each panel ($ne=2$) and $8\times 8$ quadrature points in each element ($np=8$), referred to as the $ne2np8$ grid. The average grid spacing at the Equator is approximately the same (7.5$^\circ$) for both grids. The boundaries of the elements are marked with thick black lines.}
  \label{fig:phl-gll}
 \end{figure}

\begin{figure}
 \noindent\includegraphics[width=20pc]{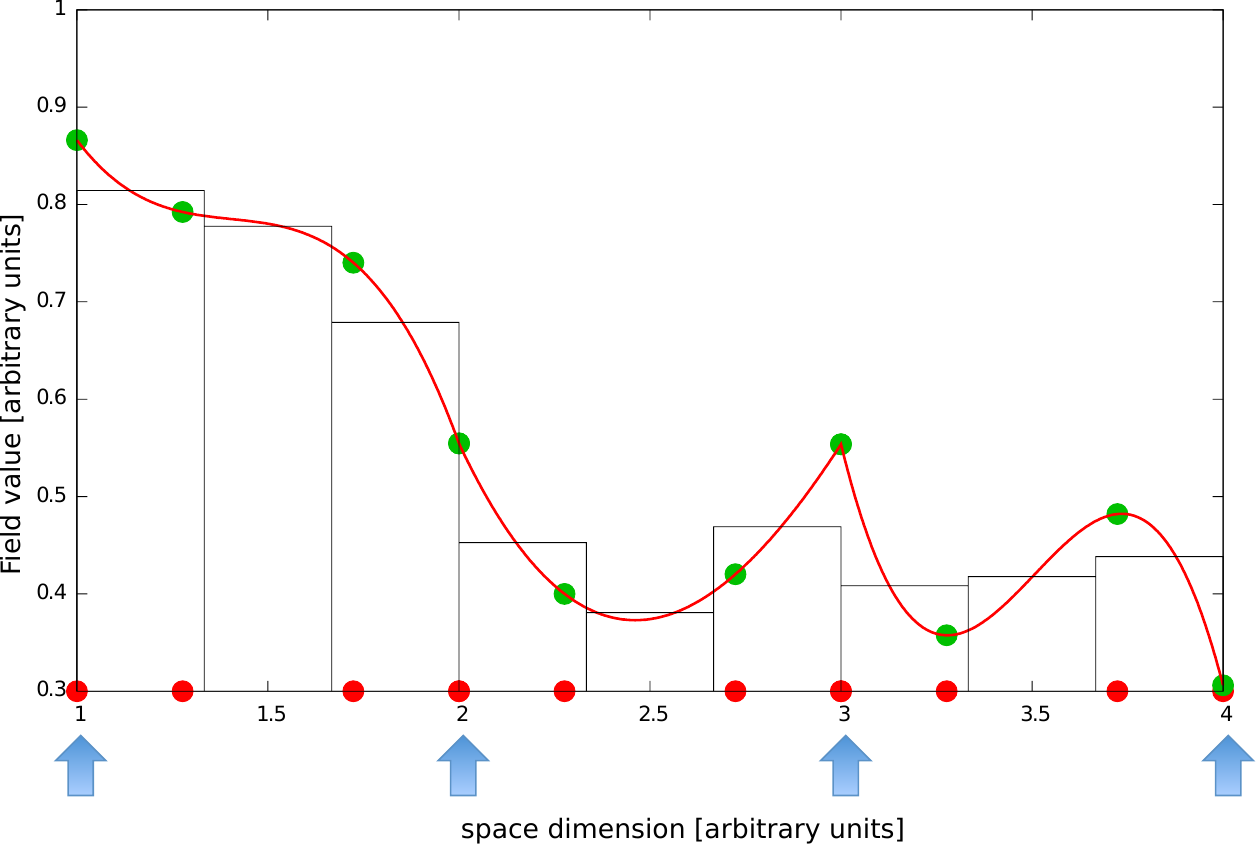}
 \caption{The figure shows three elements in one dimension. The edges of the elements are marked with blue arrows. The red curves are the degree 3 polynomials in each element and, following the \ac{CAM}-\ac{SE} algorithm, the polynomial values from each side of an element boundary are averaged. The filled green circles show the \ac{GLL} quadrature point values and the red filled circles are the location of the GLL quadrature points in each element for $np=3$. The histogram bar shows the cell averaged values on a $nc=3$ physics grid (each element has been divided into 3 equal-sized control volumes) obtained by integrating the Lagrange basis functions over the control volumes.}
  \label{fig:phl-1Dphysgrid}
 \end{figure}

Assuming that physics should be given a grid cell average value, it may be argued that it would be more consistent to integrate the basis functions within each element over quasi equal-area control volumes. From an implementation point of view it is convenient to have the control volumes sub-divide the element so that no control volume spans part of the neighboring elements. Note that the basis functions are $C^\infty$ within each element but only $C^0$ at the element boundaries. If there is a strong grid-scale forcing at the element boundary, the physics grid value may be more representative than the extrema value (see Figure~\ref{fig:phl-1Dphysgrid}). This configuration, where physics and dynamics grids are separated, is referred to as {\em{physgrid}}. Care must be taken when mapping variables to and from dynamics and physics grids so that conservation properties are not violated. Here the arbitrary high-order, conservative and consistent remapping algorithm of \citet{UT2015MWR} is used. The algorithm consists of matrices that can be pre-computed: For mapping from the dynamics to the physics grid it consists of one matrix that performs a shape-preserving, but low-order, remap and another matrix that is not shape-preserving but high-order. The algorithm has been modified such that the two matrices in each element are optimally combined (linearly) so that the method is shape-preserving and, where possible, high-order. For mapping the tendencies back to the  \ac{GLL} quadrature grid a low-order conservative and shape-preserving method is used. The mapping algorithm accommodates any order of basis functions.

The next step is the choice of physics grid resolution: Same resolution as the dynamical core, coarser or finer? \citet{Lander1997} argued, in the context of a spectral transform model, that the physical parameterizations should only be given believable scales. From linear theory it is well known that numerical methods used in the dynamical core do not represent the shortest wavelengths, e.g., the $2\Delta x$ wave, accurately. It may therefore be argued that the physical parameterizations should not be passed scales that, from linear theory, are not accurately represented. On the other hand, computing physics tendencies on a higher-resolution grid compared to the dynamical core may provide a better sampling of the atmospheric state, somewhat similar to the super-parameterization \citep{G2001JAS,GRL:GRL14999} and sub-columns concepts \citep{subcolumn,JGRD:JGRD10481,gmdd-8-5041-2015}.

In this section the consequences of separating physics and dynamics grids in \ac{CAM}-\ac{SE}, as described above, are explored. The $1^\circ$ version of  \ac{CAM}-\ac{SE} is used, i.e. the $ne30np4$ configuration ($30\times30$ elements on each panel, $ne=30$, and $4\times 4$ quadrature point, $np=4$, in each element), and the physics tendencies are computed on the \ac{GLL} grid (the grid of the dynamical core), a coarser ($1.5^\circ$) physgrid, same resolution ($1^\circ$) physgrid and finer ($0.75^\circ$) resolution physgrid.
 The four configurations are referred to as $ne30np4$, $ne30np4nc2$, $ne30np4nc3$, $ne30np4nc4$, respectively, where $nc2$ refers to a $2\times 2$ quasi equal-area physics grid in each element. Similarly for $nc3$ and $nc4$. Note that the GLL grid is the grid on which the dynamical core operates. Aqua-planet simulations \citep{NH2000ASL} are performed with \ac{CAM}4 physics \citep{CAM4} and the physics time step is the default 1800s. The reasoning behind choosing \ac{CAM}4 physics instead of the newer \ac{CAM}5 physics is that \ac{CAM}4 physics is more resolution dependent \citep[e.g.][]{BetAl2013JC,ZetAl2014JC}. \ac{CAM}4 physics is therefore expected to produce more physgrid resolution dependence than \ac{CAM}5. Simulation length is 30 months and only the last 24 months are used for analysis. The code base used is revision 65448 of \texttt{https://svn-ccsm-models.cgd.ucar.edu/cam1/branches/ physgrid}. Standard out-of-the-box namelist settings for the spectral element dynamical core were used.

\begin{figure}
 \noindent\includegraphics[width=20pc]{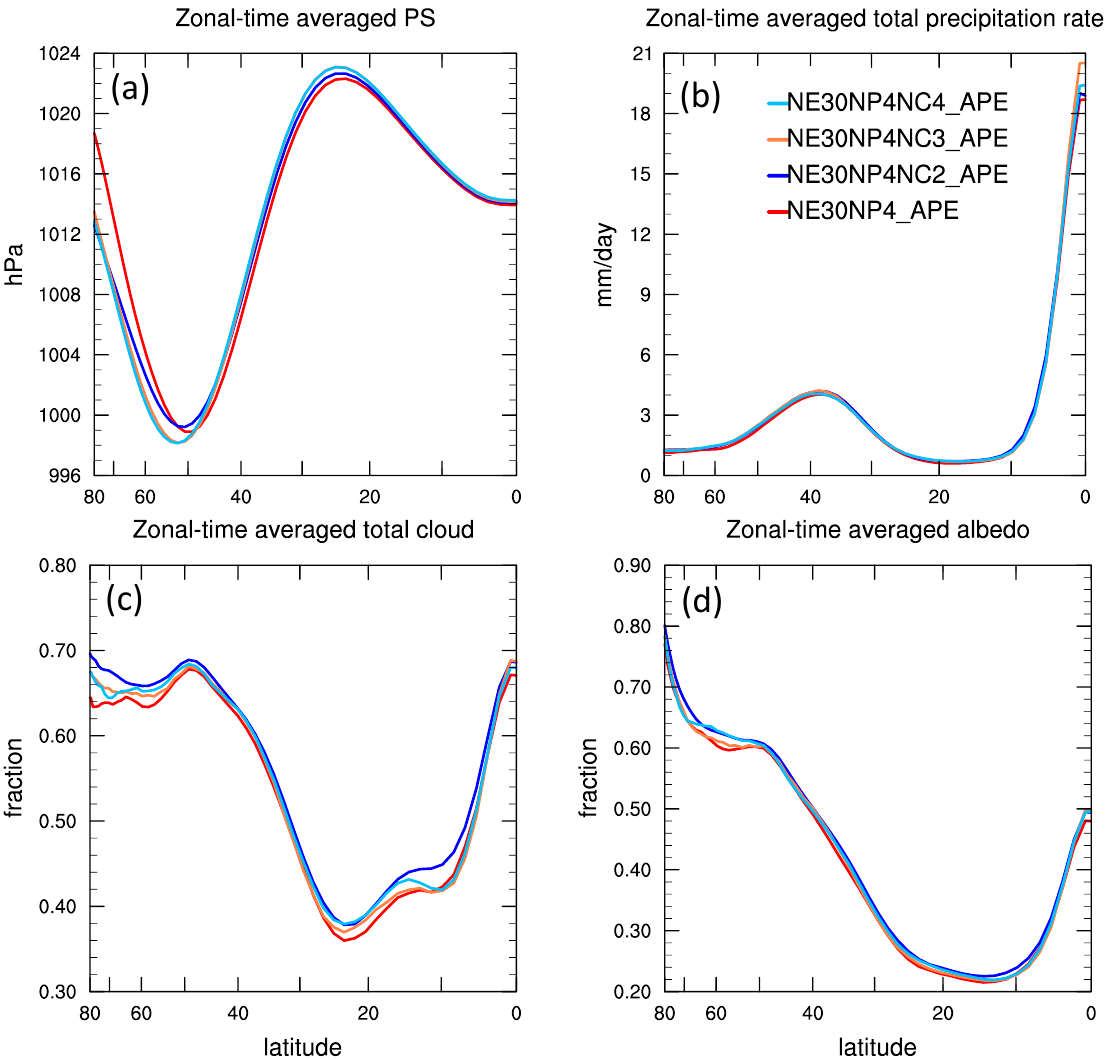}
 \caption{Zonal-time average (a) surface pressure, (b) total precipitation rate, (c) total cloud fraction, and (d) albedo as a function of latitude (from Equator to $80^\circ$N) for the different configurations of  \ac{CAM}-\ac{SE}. The data has been averaged over a period of 24 months and mapped to a $1.5^\circ$$\times$$1.5^\circ$ regular latitude-longitude grid for analysis.}
  \label{fig:phl-zonal_avg}
 \end{figure}

Figure~\ref{fig:phl-zonal_avg} shows the zonal-time average of surface pressure, total precipitation rate, total cloud fraction, and albedo as a function of latitude (from Equator to $80^\circ$N) for the different model configurations. The surface pressure field follows a slight decrease with increased physics grid resolution North of approximately 55$^\circ$N. In the simulations presented in \cite{TELA:TELA339} the surface pressure exhibits the same behavior when the model resolution was increased (see their Figure~4). Precipitation rates show little dependence on physics grid resolution except at the Equator. For total cloud fraction \citet{TELA:TELA339} observed that the fraction decreased with increasing resolution. This is observed for the physgrid $nc=2$ and $nc=3$ simulations, whereas the $nc=4$ cloud fractions are mostly bounded (in between) the $nc=2$ and $nc=4$ cloud fraction values. The same is observed for albedo. So for the zonal time-averaged fields there seems to be little dependence on physics grid resolution.

To investigate the effect of separating physics and dynamics grids on extreme events 
 the fraction of time precipitation has been binned in 1 mm d$^{-1}$ bins ranging from 0 to 120 mm d$^{-1}$ (not shown). Only a very  small dependence on physics grid resolution was observed for the high precipitation rate events.


In all, the physgrid configuration of \ac{CAM}-\ac{SE} has been demonstrated to produce aquaplanet results that are similar to the baseline (no physgrid) version. The dependence on physics grid resolution is different for different fields. The aquaplanet setup does not have stationary grid scale forcing and is only suitable for analyzing the free modes in the atmosphere. The next step is to investigate the effect of a physics grid on applications with stationary grid scale forcing (e.g., topography).  \ac{CAM}-\ac{SE} had been found to produce some noise if topography is not sufficiently smoothed \citep{gmdd-8-4623-2015}. The physgrid configuration has shown promise (not shown) in alleviating spurious grid-scale precipitation near steep orography due to the averaging over control volumes (especially near the edges of the elements). Similarly, the physgrid may improve simulations of other fields exposed to strong grid scale forcing such as photolysis driven tracers. An idealized test to investigate this has recently been developed \citep{LCLVT2015GMD}.

\subsection{Emerging challenges in \ac{PDC} with multi-scale models}
\label{sec:varres}

\begin{figure} \centering
\includegraphics[trim={0  3cm 0 0},clip,width=9cm]{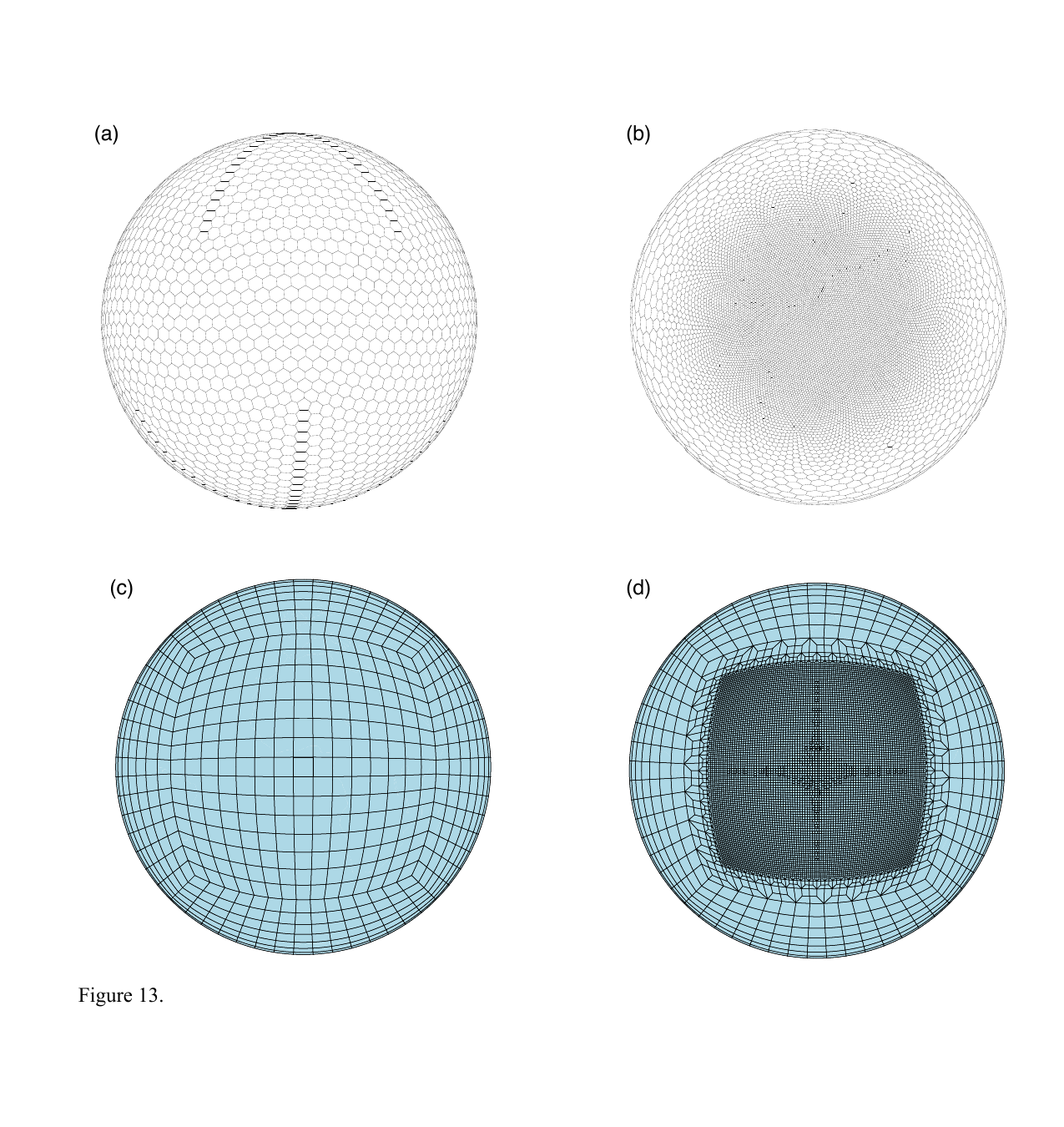}
\caption {Examples of unstructured grids with quasi-uniform and variable resolutions: (a) quasi-uniform resolution centroidal Voronoi grid used for \ac{MPAS-A}, (b) variable-resolution centroidal Voronoi grid, (c) quasi-uniform cubed sphere grid used for \ac{CAM}-\ac{SE}, and (d) variable-resolution cubed-sphere grid.
\label{Figure13} }
\end{figure}

The last decade has seen a surge of atmospheric dynamical cores formulated on quasi-uniform grids, which do not suffer from the pole problem existing in latitude-longitude grids~\citep{Williamson2007}. Their development is also driven by the need to improve scalability on massively parallel computers and by diverse model applications from weather prediction to atmospheric chemistry and climate projections. For these applications, the dynamical cores must satisfy several properties, such as conservation, compatible or mimetic properties, and accurate representation of global-to-meso-scale flows \citep{TF2010JCP,Ringler2010,Skamarock2011,StaniforthThuburn2011}. These numerical techniques, along with progress in grid generation \citep{Tomita2002,Anderson2009,Ju2011,Walko2011}, make it possible to increase grid resolution locally while maintaining a quasi-uniform resolution outside the refined domain. The associated grids are often described as unstructured because each cell is identified by a unique index and its connectivity to the neighboring cells, due to non-rectangular cell shapes and/or local coordinate system used in the numerical scheme. Unstructured grids are amenable to tiling complex geometry in irregular patterns. This is in contrast to traditional latitude-longitude grids in which each cell is regularly distributed and identified by global $x$, $y$ indexing in both physical and computational space (structured grids). Examples of unstructured grids in quasi-uniform and variable resolution configurations are shown in Figure~\ref{Figure13}. 

Local mesh refinement is also possible by stretched-grid methods on structured grids that are continuously and conformally transformed to achieve higher gridcell density over a specified region \citep{Schmidt1977,Staniforth1978}; see \citet{Fox-Rabinovitz2006} and \citet{McGregor2013} for general reviews. In this technique the number of the grid points remains the same after the transformation, so the increase in resolution over one region must be compensated by the decrease in resolution in the rest of the model domain. Recently, the stretched-grid method has been extended to unstructured grids by \citet{Uchida2016}. Here, all approaches that refine the horizontal resolutions over one or more regions on a global grid are referred to as the \ac{VR} approach. 

\begin{table*}[t]
\caption{List of references for variable-resolution modeling.}
\label{table:vr-refs}
\begin{center}
\begin{tabular}{lll}
Model Name & Grid               & References \\  \hline        
CAM-SE     & Cubed-sphere       & \cite{ZetAl2014JC,Zarzycki2014a,Zarzycki2014,Zarzycki2015,Zarzycki2015AMIP} \\
           &                    & \cite{                                                                     Rhoades2015,Huang2016} \\
HiRAM-FV3  & Cubed-sphere       &  \citet{Harris2013,Harris2014,Harris2016} \\
MPAS-A     & Centroidal Voronoi & \citet{Ringler2011,Skamarock2012,Hagos2013,Rauscher2013}      \\
           &                    & \citet{                                                 Park2013,Rauscher2014,Park2014,Sakaguchi2015,Sakaguchi2016}      \\
           &                    & \citet{                                                                                                            Fowler2016,Zhao2016}      \\
NICAM      & Icosahedral        & \citet{Tomita2008stretched,Shibuya2016,Uchida2016}   \\
OLAM       & Icosahedral        & \citet{Medvigy2008,Medvigy2010,Walko2011,Medvigy2011,Medvigy2013}
\end{tabular}
\end{center}
\end{table*}

With global \ac{VR} models, higher horizontal resolutions can be achieved in area(s) of interest while the computational burden is reduced relative to global high-resolution simulations due to coarser resolution over the remainder of the globe. The \ac{VR} approach can avoid some of the known issues in limited-area models, such as the treatment of lateral boundaries, consistency between the global and regional models, and lack of two-way interactions between the regional simulations and their driving global simulations~\citep{Wang2004}.  Idealized testing demonstrates that properly-designed numerical schemes on \ac{VR} grids can provide additional fine-scale information at the regional scale without decreasing the accuracy of the global solution~\citep{Ringler2011, Ullrich2011, Guba2014}.

The advantages and challenges of \ac{VR} weather and climate modeling have been actively studied. Table~\ref{table:vr-refs} provides a non-exhaustive list of recent studies of \ac{VR} atmospheric model simulations for interested readers. The general consensus is that \ac{VR} models can provide the benefits of high-resolution simulation inside or even outside the refined domain, such as improved orographic precipitation and snow cover~\citep{Rhoades2015}, tropical cyclones~\citep{Zarzycki2014, Zarzycki2015, Zarzycki2014a}, land cover representation \citep{Medvigy2011}, remote influence from high-resolution regions \citep{Medvigy2013, Sakaguchi2016}, and overall regional climate metrics \citep{Medvigy2010, Harris2014, Harris2016, Sakaguchi2015, Zarzycki2015AMIP, Huang2016}.  Boundary effects have also been evaluated, finding little artifacts in propagating waves throughout the variable-resolution domain~\citep{Harris2013, Hagos2013, Park2014, Zarzycki2015AMIP}. So far the most challenging issue for \ac{VR} models is related to (unphysical) sensitivity of physics parameterizations to spatial and temporal resolutions, although there are other potential challenges such as optimum topography smoothing on \ac{VR} grids [e.g., \cite{Zarzycki2015AMIP}].

The previous sections (\ref{sec:arti}, \ref{sec:time}, and \ref{sec:grey}) illustrated several examples of undesirable sensitivities of weather and climate models to temporal and spatial resolutions. Specifically, section~\ref{sec:arti} discussed the mismatch between the predefined physical process time scales and the model time steps and how the mismatch affects the interaction between convection, cloud microphysics, and resolved dynamics in the sequential-update time splitting scheme. Similar sensitivities could negatively affect \ac{VR} simulations that feature multiple resolutions within a single simulation. Indeed some studies in Table~\ref{table:vr-refs} reported non-negligible influence of resolution sensitivity on the overall quality of the \ac{VR} simulations. For example, a striking difference in precipitation appears inside and outside the high-resolution domains in aquaplanet \ac{VR} simulations using the \ac{CAM}-\ac{SE} dynamical core \citep{ZetAl2014JC} or the MPAS-A dynamical core \citep{Hagos2013,Rauscher2013,Zhao2016} with the \ac{CAM}4 and \ac{CAM}5 subgrid physics suites. In the following the effects of physics-dynamics coupling on the model sensitivity to spatial resolution in \ac{VR} modeling are briefly explored. 

Aquaplanet experiments were conducted using \ac{MPAS-A} with the \ac{CAM}4 physics as in \citet{QJ:QJ1992}, who used the Eulerian spectral model with the same \ac{CAM}4 physics. The version of \ac{MPAS-A} used is the hydrostatic model described in \citet{Park2013} and \citet{Rauscher2013}. The model was configured with three different grids: \ac{QU} $240$~km,  \ac{QU} $120$~km, and a \ac{VR} grid with $30$~km grid spacing at the center of the refined domain over the equator, transitioning to $240$~km grid spacing on the rest of the globe. There were no modifications to the \ac{CAM}4 parameterization suite, and the same tuning and configuration were used in all simulations except for the numerical diffusion coefficient, which was adjusted based on gridcell size \citep{Rauscher2013}. The same dynamics time step of 100s was used in all simulations and for each grid cell in the VR simulations. The physics time step was defined independently of the dynamics time step. For each resolution, simulations with three different ratios ($R$) of the physics time step ($\Delta t$) to the convective relaxation time scale ($\tau$) were run: $R = 1/6$ ($\Delta t = 600$~s and $\tau=3600$~s), $R=1/2$ ($\Delta t = 1800$~s and $\tau = 3600$~s), and $R = 1$ ($\Delta t = 600$~s and $\tau = 600$~s). Although different combinations of $\Delta t$ and $\tau$ can produce the same $R$ values, the analysis was limited to the aforementioned three combinations.

\begin{figure*} \centering
\includegraphics[trim={0  2cm 0 0},clip,width=\textwidth]{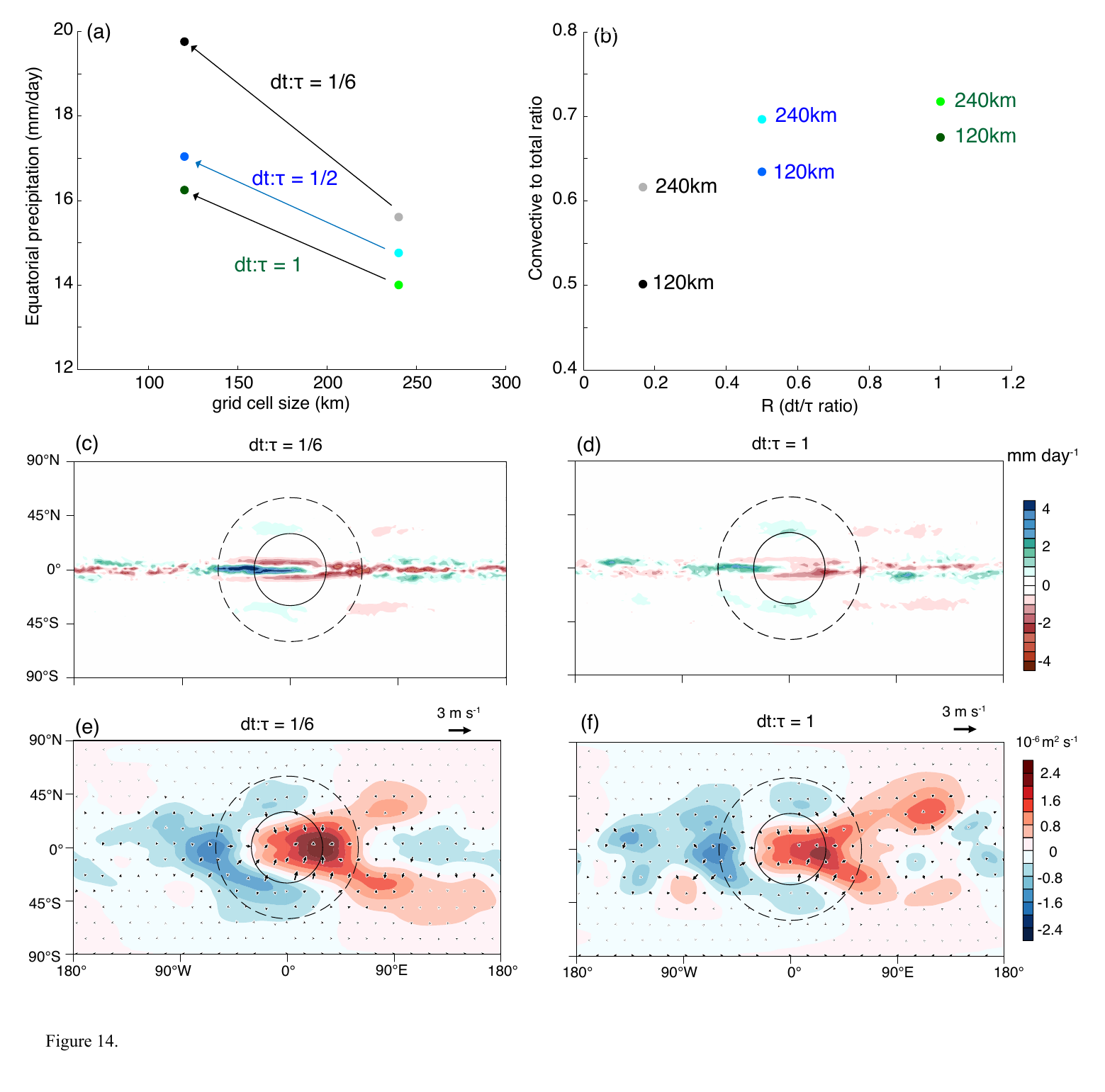}
\caption {Influence of $R$ ($\Delta t/\tau$) on the resolution sensitivity of the \ac{CAM}4 physics (precipitation) to quasi-uniform and variable resolutions using \ac{MPAS-A}. (a) Sensitivity of equatorial (+/- $2\degree$ latitude) precipitation to grid cell size (x-axis) in different values of $R$ as represented by three arrows, (b) Fraction of convective precipitation as a function of $R$ (x-axis) and grid cell size ($240$~km vs. $120$~km), (c) Zonal anomaly of precipitation in a \ac{VR} simulation with $R = 1/6$, (d) same as (c) but a VR simulation with $R = 1$, (e)  zonal anomaly of velocity potential (shading) and divergent component of wind (arrows) with $R = 1/6$, and  (f) Same as (e) but for $R = 1$. The solid and dashed circles in (c)-(f) represent the boundaries enclosing the domain with $30$~km grid and the transition to $240$~km grid domain, respectively.
\label{Figure14} }
\end{figure*}

Comparing the two \ac{QU} simulations, it is seen that as $R$ approaches unity, the sensitivity to grid spacing of the total (convective + large-scale) precipitation (Figure~\ref{Figure14}a) as well as that of the fraction of convective to total precipitation (Figure~\ref{Figure14}b) become smaller.

This dependence of the resolution-sensitivity on $R$ has a visible impact on the VR simulations. For $R$ equals $1/6$, the zonal anomaly (relative to the zonal mean) of precipitation appears on the western or downwind side of the refinement (Figure~\ref{Figure14}c), and the attendant latent heat excites a Gill-type circulation apparent in the 200-hPa velocity potential (Figure~\ref{Figure14}e, also see \citet{Hagos2013}, \citet{Rauscher2013}, and \citet{ZetAl2014JC}). Theoretically the zonal anomaly should be nearly random spatially because there is no longitude-dependent  forcings in the aquaplanet configuration. With $R = 1$, the model precipitation exhibits a substantially weaker sensitivity to the change of resolution (Figure~\ref{Figure14}d). Making $\tau$ comparable to $\Delta t$, convection is more active in removing the instability created by the resolved dynamics \citep{QJ:QJ1992}. However, the zonal anomaly is still visible and the undesirable Gill-type circulation is not entirely eliminated (Figure~\ref{Figure14}f). 

\begin{figure*} \centering
\includegraphics[trim={0  2cm 0 0},clip,width=\textwidth]{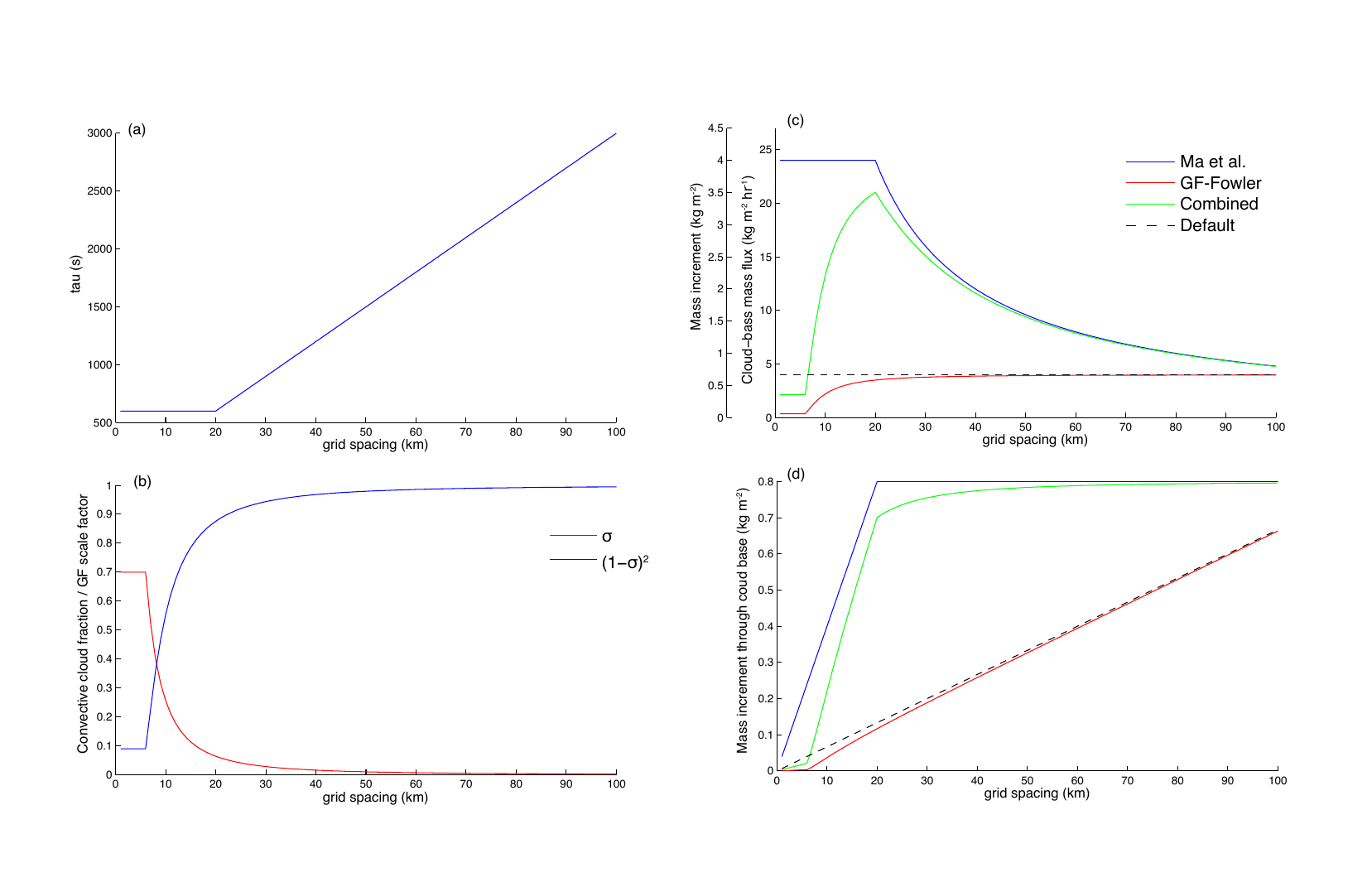}
\caption {Illustration of the \citet{Ma2014} and \citet{Fowler2016} approaches for scale-aware convection using the Zhang and McFarlane closure. In (a) $\tau$ from \citet{Ma2014} is shown as a function of grid spacing. In (b)  the fractional convective cloud cover ($\sigma$, red line) is shown  and scaling factor for cloud-base mass flux used in \citet{Fowler2016}. In (c) the cloud-base mass flux (inside y-axis) is shown based on the Zhang and McFarlene closure (equation~\ref{eq:4}) with $\textrm{CAPE} = 1000$J kg$^{-1}$ and $F = 250$ J m$^2$ kg$^{-2}$ and different modifications. The four lines represent: Dashed: the default with $\tau=3600$~s in equation~\ref{eq:6} (Default); Blue: equation~\ref{eq:6} with $\tau$ following equation~\ref{eq:4} (\citet{Ma2014}); Red: equation~\ref{eq:7} with $\tau=3600$~s (GF-Fowler); and Green: equation~\ref{eq:7} with $\tau$ as in equation~\ref{eq:4} (Combined). The outside y-axis in (c) shows the mass increment through the cloud base for $\Delta t = 600$~s (i.e., multiply each curve by $600$). In (d), the mass increment through the cloud base is shown for the same cases in (c), using $\Delta t  = 6\times \Delta x$.
\label{Figure15} }
\end{figure*}

More sophisticated modifications of convective schemes to address multiple resolutions have been proposed. From the above \ac{VR} example, one may expect more physical behavior if $\tau$ is allowed to vary with the grid spacing as opposed to using a constant value across the globe. \citet{Ma2014} and \citet{Gustafson2014} suggested a simple formulation of $\tau$ as a linear function of grid spacing:
\begin{equation}
\tau=\max\left(\tau_{min},\tau_{max}\frac{\Delta x}{\Delta x_{ref}}\right)
\label{eq:4}
\end{equation}
where $\tau_{min} = 600$~s, $\tau_{max} = 3600$~s, $\Delta x$ is the grid spacing, and $\Delta x_{ref} = 120$~km ($275$~km in \citet{Ma2014}). This function is plotted in Figure~\ref{Figure15}a. \citet{Fowler2016} tested another simple method to achieve scale-aware representation of convection in \ac{VR} \ac{MPAS-A} simulations. They used the \citet{Grell2014} (hereafter GF) convection scheme, which follows the approach originally suggested by \citet{Arakawa2011} and \citet{Arak13}. In \citet{Fowler2016}, the cloud-base mass flux is scaled by a quadratic function $(1-\sigma)^2$, with $\sigma$ being the fraction of convective cloud cover:
\begin{equation}
\sigma=\frac{0.04\pi}{A\varepsilon^2},
\label{eq:5}
\end{equation}
where $A$ is grid cell area and $\varepsilon$ is the initial entrainment rate of $7\times10^{-5} \textrm{m}^{-1}$. With this grid-size dependent scaling, convective precipitation is mostly parameterized with $>\approx 40$~km grid spacing, but the fraction of parameterized convection rapidly decreases over the $5-30$~km range, and most convection is explicitly resolved with $<5$~km grid spacing (Figure~\ref{Figure15}b). Repeating our aquaplanet simulations with the two modifications of \citet{Ma2014} and \citet{Fowler2016} is beyond the scope of this review. However, taking advantage of the simplicity of the two approaches, a heuristic, idealized analysis can be made to compare their resolution-sensitivity in the context of \ac{VR} modeling and the time-scale mismatch described in \citet{QJ:QJ1992}. 

From the closure assumption of the Zhang and McFarlane scheme (\citet{zhang:95}, hereafter ZM) used in \ac{CAM}, the cloud base mass flux ($M_b, \textrm{kg m}^{-2} \textrm{hour}^{-1}$) can be diagnosed given certain values of \ac{CAPE} and the \ac{CAPE} consumption rate per unit cloud base mass flux, $F$:

\begin{equation}
M_b=\frac{\textrm{CAPE}}{\tau F}.
\label{eq:6}
\end{equation}

Here $\textrm{CAPE}=1000$J~kg$^{-1}$ and $F=250$J~m$^{2}$kg$^{-2}$, respectively, and it is assumed that $F$ does not change substantially with resolution for a given value of \ac{CAPE}, which is reasonable based on \ac{CAM}5 simulations on a one-degree and two-degree grid with the finite-volume dynamical core. The value chosen for $F$ is higher than those typically found over tropical oceans ($F\approx 100$J~m$^2$kg$^{-2}$ for $\textrm{CAPE}\approx 1000$J~kg$^{-1}$), but convenient for showing different $M_b$ in the same plot. In the default \ac{CAM}, $\tau$ is $3600$~s. In the \citet{Ma2014} approach, the constant $\tau$ was replaced with equation~\ref{eq:4}. For the GF approach as implemented by \citet{Fowler2016}, $M_b$ can be expressed as

\begin{equation}
M_b=\left(1-\sigma\right)^2\frac{\textrm{CAPE}}{\tau F}
\label{eq:7}
\end{equation}

In Figure~\ref{Figure15}c $M_b$ is shown as a function of grid size (with the inside vertical axis).  With constant $\tau$, the default $M_b$ is constant across resolutions. With the modification following GF- Fowler (red line), $M_b$ gradually decreases with reduced grid spacing, reaching a minimum value that is specific to the implementation of \citet{Fowler2016}. This curve mimics the behavior of convection reported in their study. The \citet{Ma2014} approach exhibits a different behavior (blue line), with $M_b$ increasing with decreasing grid spacing and reaching a maximum value at a grid size that depends on the tunable parameter $\tau_{min}$. A simple combination of the two by substituting equation~\ref{eq:4} in \ref{eq:7} is also shown (green line). 

A different picture emerges when $M_b$ is multiplied by $\Delta t$ to obtain the mass increment over one time step, which is an important quantity as it represents the effectiveness of deep convection to remove instability generated over one time step. Figure~\ref{Figure15}d shows the mass increment assuming that $\Delta t$ changes with $\Delta x$, for instance by considering the \ac{CFL} stability criteria. For simplicity, $\Delta t$ is set to $ 6\times \Delta x$ (e.g., recommendation in \citet{Wang2016}). The \citet{Ma2014} approach maintains a constant mass increment independent of grid spacing for gridcells larger than the gray zone. By doing so, this approach allows deep convection to be active even as $\Delta x$ (and $\Delta t$) becomes smaller, thereby reducing the time step sensitivity elucidated by \citet{QJ:QJ1992}. The $R=1$ case in Figure~\ref{Figure14}a,b is another demonstration of the same principle of allowing deep convection to be active by making $\tau$ and $\Delta t$ comparable. On the other hand, the GF-Fowler approach produces a mass increment equal to or less than the default case. This suggests that the GF-Fowler approach will likely produce the same positive feedback and truncation-scale storms at high resolution as described in \citep{QJ:QJ1992}, if it is implemented in \ac{CAM} with the ZM scheme. This problem may be avoided by combining the GF-Fowler formulation of mass flux with equation~\ref{eq:4} (green line). Note that the closure assumption in the GF scheme is different from that in the ZM scheme and does not include a predefined convection time scale (equation~(19) in GF)), and both \citet{Gustafson2014} and \citet{Fowler2016} use the parallel-split, in which tendencies from the convection scheme do not directly affect the behavior of the cloud microphysics. Therefore the resolution-sensitivity described here is for illustrative purpose and is not directly applicable to their simulations.

In the \ac{VR} models, it is common to use a constant dynamics time step that satisfies the stability criteria of the smallest grid cell for all the gridpoints in the global domain. Generally, the physics time step ($\Delta t$) is also fixed in relation to the dynamics time step. In this case the mass increment over $\Delta t$ behaves in the same way as $M_b$ (i.e., as in Figure~\ref{Figure15}c), because $M_b$ is multiplied by the same constant $\Delta t$ regardless of the grid spacing. The outside vertical axis of Figure~\ref{Figure15}c shows the mass increment for a constant $\Delta t$ of $600$~s. Based on this plot, the behavior of the GF-Fowler approach seems to be more desirable for \ac{VR} models with a constant $\Delta t$ and the \ac{CAM}'s sequential-update splitting. The modification by \citet{Ma2014} (and the result in Figure~\ref{Figure14}d,f with $R=1$) would introduce larger mass increment from the parameterized convection for smaller grid spacing even beyond the gray zone, which is asymptotically erroneous. This artifact arises from the implicit assumption that $\Delta t$ and $\Delta x$ vary together, but this assumption breaks down in \ac{VR} models. This simple analysis illustrates the dependence of scale-aware convection representation on the Physics-Dynamics coupling, such as the time splitting method or the co-variation between $\Delta t$ and $\Delta x$, which has not been elucidated in the \ac{VR} modeling framework.  

It is worth noting that global simulations using the newer \ac{CAM}5 parameterization suite shows improvement in some of these aspects, particularly with respect to cloud fraction and precipitation scaling in \ac{VR} simulations \citep{ZetAl2014JC,Zarzycki2015AMIP,Zhao2016}. Both \ac{CAM}4 and \ac{CAM}5 use the same deep convective scheme; therefore aspects of the scale sensitivity seen with \ac{CAM}4 may not be purely driven by the lack of scale-awareness such as related to the convective timescale. In aquaplanet simulations, \citet{ZetAl2014JC} found a smaller scale sensitive precipitation response with increasing resolution when using the \ac{CAM}5 parameterization suite versus \ac{CAM}4, a result that was replicated using more complex \ac{AMIP}-style simulations with regional refinement \citep{Zarzycki2015AMIP}. In these cases, smaller anomalies in total precipitation with \ac{CAM}5 are noted across grid spacings, which reduce the magnitude of spurious, physics-induced circulations as described above. Additionally, \citet{OBrien2013} postulated that the large improvement in cloud fraction scaling is actually dominated by \ac{CAM}5's new microphysical parameterizations. However, these results underscore the need to understand the complex relationships between the multitudes of components within parameterization suites that continue to grow in complexity.

In summary, examples illustrating process-coupling issues unique to \ac{VR} models were presented. The example using \ac{MPAS-A}-\ac{CAM}4 demonstrated the relevance of time step/time-scale interaction \citep{QJ:QJ1992} to the \ac{VR} framework. The analysis using the closure of the ZM scheme is admittedly idealistic, but illustrative of the subtle differences between two simple scale-aware approaches for representing convection (i.e., \citet{Ma2014} and \citet{Fowler2016}) and provided some testable hypotheses. In particular, a common practice in \ac{VR} models to use the same physics time step across all gridpoints (regardless of their size) may not produce the desired effects of scale-aware convection parameterizations, depending on the type of process splitting. Previous studies also noted that scale-awareness in other parameterizations plays a role in determining the overall resolution-sensitivity and coupling to the dynamics. Based on the review and analysis, it is understood that novel approaches for understanding the interaction between resolved dynamics and subgrid parameterizations are required to fully exploit emerging themes in model development. For instance, a wide spectrum of scale-aware parameterizations (cf. section~\ref{sec:grey}) as well as a variety of splitting methods (cf. section~\ref{sec:arti}) should all be analyzed and evaluated in the \ac{VR} framework. Having $\Delta t$ dependent on the gridcell size may be one pathway to remove one dimension of the resolution-sensitivity specific to \ac{VR} models so that scale-aware parameterizations tested for quasi-uniform resolution models would perform equally in \ac{VR} models, although this could be technically challenging. An alternative pathway is to use a separate, uniform-resolution physics grid (cf. section~\ref{sec:emerging_fe}) for calculation of tendencies from physics parameterizations. Such a separate physics grid was adopted by \citet{Fox-Rabinovitz2001} for their stretched-grid model. Addressing these physics-dynamics and process coupling issues, in addition to other fundamental challenges with respect to the current generation of numerical schemes and physics parameterizations, will be necessary to take full advantage of the capability of \ac{VR} models.

\section{ Conclusions and outlook }
\label{sec:conc}

The wide span of issues related to \ac{PDC} in geophysical models has been presented. It is apparent that under the surface of all of these remain many unanswered questions.

The current efforts are disperse. The historical review in section~\ref{sec:hist} has highlighted that research in this area has been ongoing for some time by various groups focusing on individual aspects. However, this research does not always apply in a direct fashion to full models, partly due to the highly complex and cross disciplinary nature of the topic.

Probably the most sensitive parameter is the time step. Section~\ref{sec:arti} has shown that there is ample evidence in the commonly used models relating to a time step sensitivity. The cause and effect, however, is not as clearly identifiable. As section~\ref{sec:time} has shown, convergence of the whole model run with a reduction of the time step is far from trivial to achieve and analyze.
This is where the value of reduced models comes in. As has been shown above, there are different ways in which reduction can be achieved. One option is to reduce the equation set, as in section~\ref{sec:simplified_eq}, which then renders the generation of a reference solution more straight forward and allows for more rigorous mathematical analysis. Another option, as discussed in section~\ref{sec:simple}, is to reduce the complexity of the \ac{GCM}. Obviously the balance has to be right. Oversimplification does not challenge the coupling as the real model would, overly complex setups make the analysis intractable. 

The topic has then been taken to the next level of complexity when discussing the challenge of coupling two models to each other, as in the Atmosphere-Ocean case. In section~\ref{sec:intra} this has been described theoretically and reference been made to potential analysis methods.

In the background of all of the modeling and coupling activities naturally exists one fundamental common set of laws. Thermodynamics, the subject of section~\ref{sec:thermod}. Naturally one would think that all models should obey the laws of thermodynamics. However, due to different formulations and the at times more pragmatic attitudes towards the representation of sub grid process and coupling to or coupling of models, the conformance with the laws of thermodynamics may not be as strict as would be desirable. 

One could argue that this could be corrected readily. This however overlooks at least one crucial aspect: The wide range of parameters that exists and the models are subjected to. One of these aspects is resolution, for example. Models are run at different resolutions for different purposes or resolution passes through regions where one formulation ceases to be appropriate (parameterization) and the processes are resolved by the fundamental equations of motion (resolved convection). This problem has been discussed and illustrated in section~\ref{sec:grey}. 

As well as the sensitivity for the subject, the individual models and modeling approaches evolve with time. As pointed out by the last section, the emergence of variable resolution models and  high-order Galerkin methods poses new challenges and yet another level of complexity for the coupling question. For example the gray zone now coexists with regions where the parameterization is well justified and regions where the model resolves processes, in the same model run. Not only is there the question of when in time to evaluate the parameterizations, or how to apply their forcings in time, there is also a choice of what the parameterizations should see, the element averaged values or node integration point values and how their forcings should then be applied.

Perhaps the hardest yet most important is how to analyze model performance in terms of the different schemes and hence determine which scheme is better. In order to achieve this, however, the analysis has to touch every one of the above aspects, in order to be relevant for the final model outcome. This is a challenge which the community has to tackle as a whole.
The generation of standardized hierarchies of setups with their respective established reference solutions should aid in the translation from improved components into improved full model runs and enable insightful evaluation of changes made to the coupling. 

Overall, decisions have to be made. Decisions demand guidance, objectively and systematically. This is an activity that impacts all of the modeling community, from developers to users. Due to its complexity it also has to be tackled by the community as a whole. The authors hope that this article will seed this development and provide a basis for this decision-making process and that the PDC workshop series (2014 in Ensenada, BC, Mexico, 2016 in Richland, WA, USA and 2018 in Reading, UK (tbc) - refer to \texttt{http://pdc.cicese.mx} for the latest information and material from the previous workshops) will provide a platform for this.

\acknowledgments
The National Center for Atmospheric Research is sponsored by the National Science Foundation. Peter Hjort Lauritzen (NCAR) would like to acknowledge the many discussions on high-order methods with Ram D. Nair (NCAR) and thank Paul A. Ullrich (UCDavis) for help on implementing his remapping method. The physgrid work would not have been possible without the support of Steve Goldhaber (NCAR) and Mark~A.~Taylor (SNL) who were partially funded by the Department of Energy Office of Biological and Environmental Research, work package 12-015334 ``Multiscale Methods for Accurate, Efficient, and Scale-Aware Models of the Earth System''.
Hui Wan was partially supported by the Linus Pauling Distinguished Postdoctoral Fellowship of the Pacific Northwest National Laboratory (PNNL) through the Laboratory Directed Research and Development Program.
Hui Wan, Peter Caldwell, and Phil Rasch acknowledge support from the DOE Office of Science as part of the Scientific Discovery through Advanced Computing (SciDAC) Program.
Christiane Jablonowski and Diana Thatcher (University of Michigan) were supported by the DOE Office of Science grants DE-SC0006684 and DE-SC0003990.
Koichi Sakaguchi and L. Ruby Leung were supported by the U.S. Department
of Energy (DOE) Office of Science Biological and Environmental Research as
part of the Regional and Global Climate Modeling program, and used the
computational resources from the National Energy Research Scientific
Computing Center (NERSC), a DOE User Facility supported by the Office of
Science under Contract DE-AC02-05CH11231, and the PNNL Institutional Computing. Koichi Sakaguchi would
like to thank Drs. Sara Rauscher (Univ. of Delaware), Chun Zhao (PNNL), and
Jin-Ho Yoon (PNNL) for their help on the MPAS simulations, and Samson Hagos
(PNNL) for helpful discussions. PNNL is operated for DOE by Battelle
Memorial Institute under contract DE-AC05-76RL01830.
F.  Lemari\'{e} and E. Blayo appreciate support from the french national research agency through contract ANR-14-CE23-0010 (HEAT).
The authors acknowledge the contribution to the field made by Jean-Fran\c{c}ois Geleyn during his life time. The references to his work just in this paper are a clear testament of his broad and in depth contributions. Unfortunately health concerns prohibited him from delivering his keynote lecture at PDC14. Memoratus in aeternum. *1950  $\dagger$8 January 2015.

\bibliographystyle{agufull08}

\end{document}